\newcommand*{\ATLASLATEXPATH}{latex/}
\documentclass[cernpreprint, UKenglish, txfonts,texlive=2011]{\ATLASLATEXPATH atlasdoc}

\usepackage{\ATLASLATEXPATH atlascontribute}

\usepackage{\ATLASLATEXPATH atlasphysics}
\usepackage[subfigure=true,block=none]{\ATLASLATEXPATH atlaspackage}

\usepackage{\ATLASLATEXPATH atlasbiblatex}

\usepackage{latex/main-defs}

\makeatletter
\ifthenelse{\APKG@texlive < 2012}{
  \sisetup{load-configurations=version-1}
}{
  \sisetup{version-1-compatibility}
}
\makeatother

\newcommand{\tabitem}{~~\llap{\textbullet}~~}

\usepackage{multirow}

\newcommand{\GEV}{\GeV}
\newcommand{\TEV}{\TeV}

\DeclareSIUnit[number-unit-product = {}] \TEV{
\ifmmode {\mathrm{\ Te\kern -0.1em V}}\else
\textrm{Te\kern -0.1em V}\fi}

\DeclareSIUnit[number-unit-product = {}] \GEV{
\ifmmode {\mathrm{\ Ge\kern -0.1em V}}\else
\textrm{Ge\kern -0.1em V}\fi}

\graphicspath{{logos/}{figures/}}
\usepackage{placeins}

\hypersetup{pdftitle={\AtlasTitleText},pdfauthor={The ATLAS Collaboration}}

\AtlasTitle{Search for the production of single vector-like and excited quarks in the $\Wt$ final state in $pp$ collisions at
$\sqrt{s} = \SI{8}{\TeV}$ with the ATLAS detector}

\PreprintIdNumber{CERN-PH-EP-2015-235}

\AtlasJournal{JHEP}

\AtlasAbstract{ %
A search for vector-like quarks and excited quarks in events containing a top quark and a $W$~boson in the final state is reported here.
The search is based on $\SI{20.3}{\ifb}$ of proton--proton collision data taken at the LHC at a centre-of-mass energy of $\SI{8}{\TEV}$ recorded by the ATLAS detector.
Events with one or two leptons, and one, two or three jets are selected with the additional requirement that at least one jet contains a $b$-quark.
Single-lepton events are also required to contain at least one large-radius jet from the hadronic decay of a high-\pT\ $W$ boson or a top quark.
No significant excess over the expected background is observed and upper limits on the cross-section times branching
ratio for different vector-like quark and excited-quark model masses are derived. For the excited-quark production and decay to $\Wt$ with unit couplings, quarks with masses below \SI{1500}{\GEV} are excluded and coupling-dependent limits are set.

}

\begin{document}

\maketitle
\tableofcontents

\section{Introduction}
\label{sec:intro}

The number of quark generations known within the Standard Model (SM) is three and the existence of additional heavy
quarks similar to those in the SM is strongly constrained by the discovery of the Higgs boson at the Large Hadron Collider (LHC)~\cite{Eberhardt:2012gv,HIGG-2012-27,CMS-HIG-12-028}. Additional quarks that have non-SM Higgs couplings, in particular vector-like quarks (VLQs), remain popular, especially in models which address the naturalness question~\cite{Holdom:2009rf,Alok:2010zj}.
New quarks can have right and left handed couplings to the W boson and are color triplets~\cite{AguilarSaavedra:2009es}.
VLQs appear in several extensions of the SM, such as extra dimensions~\cite{Randall:1999ee}, supersymmetry~\cite{Martin:2009bg}, composite
Higgs~\cite{Kaplan:1983sm,Vignaroli:2012sf} and little Higgs~\cite{Schmaltz:2005ky} models, in which they cancel top-quark loop
contributions to the Higgs mass~\cite{Xiao:2014kba}. Depending on the model, VLQs can be realized in different multiplets, such as in singlets, doublets and triplets~\cite{delAguila:2000rc}.

\begin{figure}[htbp]
  \centering
  \subfigure[]{
  \label{fig:BprimeFeynman}
    \includegraphics[width=0.48\textwidth]{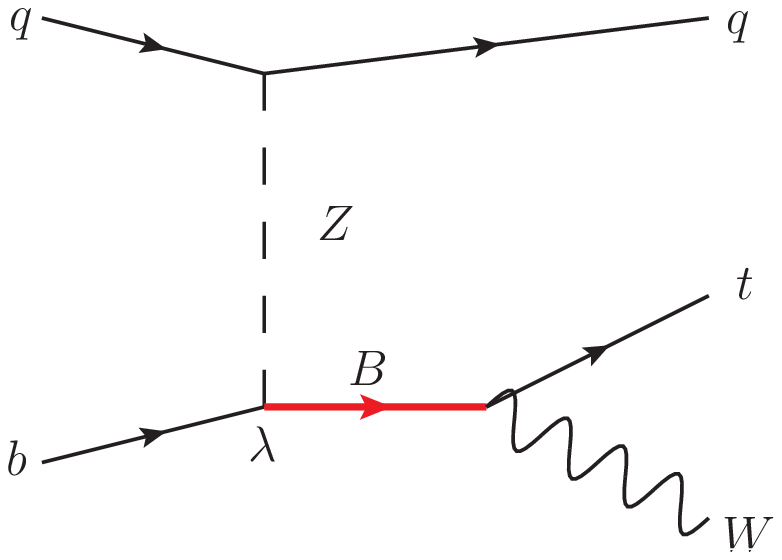}
  }
  \subfigure[]{
  \label{fig:BpFeynman}
    \includegraphics[width=0.48\textwidth]{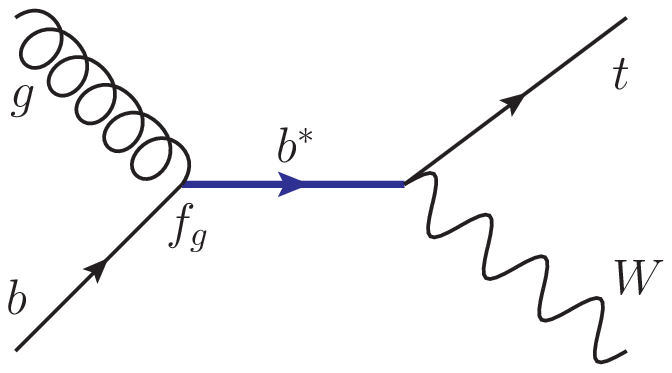}
  }
  \caption{Leading-order Feynman diagrams for the production and decay of~\subref{fig:BprimeFeynman} a single \Bprime\ together with a light quark
  and \subref{fig:BpFeynman} a single \bstar. \label{fig:SignalFeynman}}
\end{figure}

This paper describes a search for singly produced vector-like quarks with charge $\pm\frac{1}{3}$ of the elementary charge, $e$.
Two models are considered: the single production of a VLQ via the $t$-channel exchange of a $Z$~boson in a composite Higgs model~\cite{Vignaroli:2012sf,Vignaroli:2012si} (called $\Bprime$), and the single production of a VLQ that also has excited-quark couplings~\cite{RS:2012} (called $\bstar$).
Only final states in which the \Bprime\ or \bstar\ decay into $\Wt$ are considered.
The corresponding leading-order (LO) Feynman diagrams for these processes are shown in Figure~\ref{fig:SignalFeynman}.

The cross-section for singlet $\Bprime$ production is proportional to the square of the $bZ\Bprime$ coupling strength $\lambda$.
The production via Higgs-boson exchange is also possible, although the $Z$-induced process is dominant.
The $\Bprime$ can decay into $Zb$, $Hb$ and $\Wt$ with the branching ratios given by the VLQ couplings\footnote{At high singlet $\Bprime$
masses they are roughly 1:1:2~\cite{AguilarSaavedra:2009es}.} for singlet $\Bprime$.
The light quark in the final state gives rise to a forward jet. The cross-section for singlet $\Bprime$ production with subsequent decay to $\Wt$ has been calculated in the TS-10 model~\cite{Vignaroli:2012sf,Vignaroli:2012si}, a four-dimensional version of a model with composite fermions in a 10~representation of SO(5).
The cross-section is given in Table~\ref{tab:Bprimexs} for two values of the coupling parameter $\lambda$, for which the $2 \times 2$ mass mixing matrix of the $b$-quark and the $\Bprime$ has been diagonalised. 
The largest value of $\lambda$ for which the $\Bprime$ decay width is still smaller than the experimental mass resolution is $\lambda =3$.
A top-quark mass of $\SI{172.5}{\GEV}$ is assumed throughout.

\begin{table}[!h!tbp]
  \centering
  
  \begin{tabular}{S[table-format=4.0] S S}
  \toprule
  {$\mBprime$ [\si{\GEV}]} & \multicolumn{2}{c}{$\sigma  \times  \mathrm{BR}(\Bprime\rightarrow \Wt)$ [fb]} \\
                           & {$\lambda=2$}  & {$\lambda=3$}  \\ 
  \midrule
  400  & 710 & {--- } \\ 
  600  & 220 & 250    \\ 
  800  & 52  & 97     \\ 
  1000 & 15  & 30     \\ 
  1200 & 4.8 & 10.2   \\ 
  1400 & 1.7 & 3.6    \\ 
  
  \bottomrule
\end{tabular}
\caption{Cross-section times branching ratio for $pp \rightarrow \Bprime q \rightarrow \Wt$ for different $\Bprime$ masses and coupling values $\lambda$ at a centre-of-mass energy of $\sqrt{s} = \SI{8}{\TEV}$~\cite{Vignaroli:2012sf,Vignaroli:2012si}. 
For $ \lambda \cdot v / \sqrt{2} > m_B - m_b$ with the vacuum-expectation value $v$ and the mass of the $b$-quark $m_b$ one gets an unphysical $b$-quark mass. This is denoted in the table by "{---}".
}
\label{tab:Bprimexs}
\end{table}

In addition to the $Zb$, $Hb$ and $\Wt$ couplings, the $\bstar$ also has a chromomagnetic coupling $f_g$ to a gluon and a $b$-quark~\cite{RS:2012,Chivukula:2013kw,Hill:2002ap}, making it both a vector-like quark and an excited quark~\cite{Baur:1987ga}.
Complete models of VLQs usually contain other particles and interactions which result in an effective chromomagnetic coupling.
Examples of such models are technicolour~\cite{Weinberg:1975gm,Susskind:1978ms}, topcolour~\cite{Hill:1980sq,Chivukula:1998wd}, extra dimension models~\cite{Cheung:2007bu,Fitzpatrick:2007sa} or models with a heavy partner of the gluon~\cite{Bini:2011zb}, which gives rise to the effective $gb$ coupling.
The strength of the coupling is assumed to be $\fg=1$ in this paper unless otherwise indicated. 
The $\bstar$ in the VLQ case has $\fL=\fR=1$, but it is also allowed to have purely left-handed ($\fL=1$, $\fR=0$) or purely right-handed ($\fL=0$, $\fR=1$) couplings to the top quark.
The cross-sections times branching ratios for the production of $\bstar$ with decay to $\Wt$ for these three coupling scenarios are given in Table~\ref{tab:bstarxs}.
For purely left-handed couplings the cross-sections are the same as those for purely right-handed couplings. However, for VLQ-like couplings the cross-sections are not exactly the sum of the individual contributions since setting both couplings to 1 modifies the $\bstar$ decay branching ratios. 
In the following, $\fL=\fR=1$ is assumed unless stated otherwise.

\begin{table}[htbp]
\centering
  \begin{tabular}{S[table-format=4.0] S S}
  \toprule
$\mbstar$ [\si{\GEV}] & \multicolumn{2}{c}{$\sigma \times \mathrm{BR}(\bstar\rightarrow \Wt)$ [fb]} \\
   & {$f_{\textrm{L(R)}}=1$, $f_{\textrm{R(L)}}=0$} & {$\fL=\fR=1$} \\
\midrule
 400 & $ \num{115e3}$  & $\num{196e3}$ \\
 600 & $\num{18.3e3}$   & $\num{35e3}$  \\
 800 &  $\num{3.9e3}$  & $\num{7.5e3}$ \\
1000 & $\num{1.0e3}$   & $\num{2.0e3}$ \\
1200 & 310  & 610 \\
1400 & 110  & 210\\
1800 & 16 & 30\\

\bottomrule
\end{tabular}
\caption{Cross-section times branching ratio for $\bstar \rightarrow \Wt$ for different $\bstar$ masses and $\bstar \Wt$ couplings~\cite{RS:2012} at a centre-of-mass energy $\sqrt{s} = \SI{8}{\TEV}$. Here $\fg=1$ is assumed.}
\label{tab:bstarxs} 

\end{table}

After the decay of $\Bprime$/$\bstar$ into $\Wt$, the top~quark decays into a $b$-quark and another $W$~boson. 
At least one of the $W$~bosons is required to decay leptonically (to an electron or muon, and the corresponding neutrino). 
Events are separated into dilepton and single-lepton signatures. 
A hadronically decaying $W$~boson or top~quark is identified by clustering its decay products into a single jet since for high $\Bprime$/$\bstar$ masses the $W$~boson and top~quark are boosted. 
The clustering requirement is the main change in the analysis strategy for the single-lepton channel compared to the search for single-$\bstar$ production in the complete \SI{7}{\TEV} LHC dataset collected by ATLAS~\cite{TOPQ-2012-09}.
It increases the sensitivity to high $\Bprime$ and $\bstar$ masses, where the top~quark and $W$~boson have large transverse
momentum.\\
Searches for a VLQ with charge $\pm\frac{1}{3}e$ have also been performed by ATLAS in final states with $Z$-bosons using data at $\sqrt{s}=\SI{8}{\TEV}$, and exploiting both pair production and single production and resulted in limits in the range \SIrange[range-units = single]{685}{755}{\GEV}~\cite{EXOT-2013-17}.
Searches for pair production of VLQs, assuming strong interactions similar to those in the SM, have also been performed by
ATLAS~\cite{Aad:2015kqa,Aad:2015mba,Aad:2015gdg} in data at $\sqrt{s}=\SI{8}{\TEV}$ and by CMS~\cite{CMS-B2G-12-004,CMS-EXO-11-036} at
${\sqrt{s}=\SI{7}{\TEV}}$, and resulted in limits in the ranges \SIrange[range-units = single]{760}{900}{\GEV}~and \SIrange[range-units = single]{625}{675}{\GEV}, respectively.
Recently, the CMS collaboration set lower limits on pair-produced vector-like $B$~quark mass in the range \SIrange[range-units = single]{740}{900}{\GEV}, for
different combinations of the $B$~quark branching fractions~\cite{CMS-B2G-13-006} and on left-handed, right-handed, and vector-like
$\bstar$~quark masses decaying to $\Wt$ in the range \SIrange[range-units = single]{1390}{1530}{\GEV}~\cite{CMS-B2G-14-005}, using data at
$\sqrt{s}=\SI{8}{\TEV}$.
A search for $\bstar$ has been performed by ATLAS with the full \SI{7}{\TEV} LHC dataset and a limit on the $\bstar$ mass for couplings $\fg=\fL=\fR=1$ was set at \SI{1.03}{\TEV}~\cite{TOPQ-2012-09}.
All limits are given at 95\% credibility level (CL). 
This search for the single production of $\Bprime \to \Wt$ is the first for a $\Bprime$ in this final state and using the novelty approach for boosted event topologies.
While searches for pair production currently dominate the limits, single production becomes more competitive for higher quark masses, due to the reduced kinematic phase-space constraints and the larger dataset.

In the analysis presented here, six different single-lepton and dilepton signal regions (SR) are defined, which were optimised to maximise the expected significance for the $\Bprime/\bstar$ models considered.
This analysis is also complementary to the search of dijet mass resonances in ATLAS with data at $\sqrt{s}=\SI{8}{\TEV}$, which probes the $qg$ final state~\cite{EXOT-2013-11}. 
The main processes contributing to the background after applying all selection cuts are from top-quark pair ($\ttbar$) production and single top-quark production in association with a $W$~boson. Other background contributions are from $W$- or $Z$-boson production in association with jets.
Smaller contributions arise from diboson production processes and processes where a jet is misidentified as a lepton.
To estimate these SM backgrounds in a consistent and robust fashion, corresponding control regions (CR) are defined for each of the signal regions. 
They are chosen to be non-overlapping with the SR selections in order to provide independent data samples enriched in particular background sources. 
The shape of the discriminating variable, the invariant mass for the single-lepton channel and transverse mass for the dilepton channel, is used in a binned likelihood fit to test for the presence of a signal.
The systematic uncertainties and the MC statistical uncertainties on the expected values are included in the fit as nuisance parameters.
These additional degrees of freedom allow the modelling of the backgrounds to be improved based on data, increasing the sensitivity to the signal.
Correlations of a given nuisance parameter across the various regions, between the various backgrounds, and possibly the signal, are also taken into account. 
A background-only fit is used to determine the compatibility of the observed event yield in each SR with the corresponding SM background expectation. 
The improved \textit{post-fit} model resulting from this fit is used throughout this paper for the presentation of control distributions.
The observed and expected upper limits at $95\%$ CL on the number of events from VLQ phenomena for each signal region are derived using a Bayesian approach.

\section{ATLAS detector}
\label{sec:detector}

The ATLAS experiment~\cite{PERF-2007-01} at the LHC is a multi-purpose particle detector
with a forward-backward symmetric cylindrical geometry and a near $4\pi$ coverage in
solid angle.\footnote{ATLAS uses a right-handed coordinate system with its origin at the
nominal interaction point (IP) in the centre of the detector and the $z$-axis along the beam
pipe. The $x$-axis points from the IP to the centre of the LHC ring, and the $y$-axis points
upwards. Cylindrical coordinates $(r,\phi)$ are used in the transverse plane, $\phi$ being the
azimuthal angle around the $z$-axis. The pseudorapidity is defined in terms of the polar angle
$\theta$ as $\eta = -\ln \tan(\theta/2)$. Angular distance is measured in units of
$\Delta R \equiv \sqrt{(\Delta\eta)^{2} + (\Delta\phi)^{2}}$.}
It consists of an inner tracking detector surrounded by a thin superconducting solenoid
providing a \SI{2}{\tesla} axial magnetic field, electromagnetic (EM) and hadron calorimeters, and a
muon spectrometer. The inner tracking detector covers the pseudorapidity range $|\eta| < 2.5$.
It consists of silicon pixel, silicon microstrip, and transition radiation tracking detectors.
Lead/liquid-argon (LAr) sampling calorimeters provide electromagnetic energy measurements
with high granularity. A hadronic (iron/scintillator-tile) calorimeter covers the central
pseudorapidity range ($|\eta| < 1.7$).
The end-cap and forward regions are instrumented with LAr calo\-rimeters for both the EM and hadronic
energy measurements up to $|\eta| = 4.9$.
The muon spectrometer surrounds the calorimeters and is based on three large air-core toroid
superconducting magnets with eight coils each. 
It includes a system of precision tracking chambers up to $|\eta| = 2.7$ and fast detectors for triggering at $|\eta| < 2.4$.
A three-level trigger system is used to select events of interest.
The first-level trigger is implemented in hardware and uses a subset of the detector information, while the other two levels are software-based, and the last trigger level uses
the full detector information.

\section{Data and simulated samples}
\label{sec:DataMC}

The dataset used for this analysis was collected in 2012 by the ATLAS detector at the LHC, and corresponds to an integrated luminosity of \SI{20.3}{\ifb} of
proton--proton collisions at a centre-of-mass energy $\sqrt{s}=\SI{8}{\TEV}$.
Events are required to have passed at least a single-electron or single-muon trigger.
The electron and muon triggers impose a transverse momentum ($\pT$) threshold of \SI{24}{\GEV} along with isolation requirements on the lepton.
To recover efficiency for higher-$\pT$ leptons, the isolated-lepton triggers are complemented by triggers without isolation requirements but with a higher $\pT$ threshold of \SI{60}{\GEV} (\SI{36}{\GEV}) for electrons (muons).

Events are accepted if they contain at least one reconstructed primary vertex (PV). 
A reconstructed PV candidate is required to have at least five associated tracks of $\pT > \SI{400}{MeV}$, consistent with originating from the beam collision region in the $x$–$y$ plane.
The PV is chosen as the vertex candidate with the largest sum of the squared transverse momenta of its associated tracks among all candidates.

The $\Bprime$ signal is simulated with {\scshape Protos} v2.2~\cite{AguilarSaavedra:2009es}, interfaced to {\scshape Pythia} v6.4~\cite{Sjostrand:2006za} for hadronisation and underlying events. 
The MSTW2008LO68CL~\cite{Martin:2009iq,Martin:2009bu} parton distribution function (PDF) set is used.
The production kinematics and decay properties do not depend on the coupling $\lambda$.
The $\bstar$ signal is simulated at LO in QCD with the matrix-element generator {\scshape MadGraph5} v1.5.12~\cite{Alwall:2011uj}
and interfaced to {\scshape Pythia} v8.175~\cite{Sjostrand:2007gs} for hadronisation. The MSTW2008LO6CL PDF set is used.
The couplings are set to $\fg=\fL=1$, $\fR=0$ for the left-handed samples, and $\fg=\fR=1$, $\fL=0$ for the right-handed ones.
For both signal models, samples are generated with the mass of the new quark set to values from \SI{400}{\GEV} to \SI{1800}{\GEV}. 
The {\scshape Protos} $\Bprime$ sample is generated by the diagram $gq \to Bbq$ and the factorisation and renormalisation scales are set dynamically to the square of the momentum transfer of the virtual $Z$~boson for the light quark line and to the $\pT$ of the extra $b$-quark for the gluon splitting line. 
In the {\scshape MadGraph5} samples, the scales are set to the mass of the generated resonance particle.
Since for most of the analysis presented here it is assumed that $\fL=\fR=1$, left- and right-handed samples are added together and scaled
to the appropriate theory cross-section times branching fraction.

Top-quark pair and single top-quark events in the $t$, $s$ and $\Wt$ channels are simulated using the next-to-leading-order (NLO) generator {\scshape Powheg-Box} v1\_r2129, v1\_r1556 and v1\_r2092, respectively~\cite{Alioli:2010xd,Frixione:2007nw}.
The CT10~\cite{Lai:2010vv} PDF set is used and {\scshape Pythia} v6.4~\cite{Sjostrand:2006za} performs the hadronisation.
The top-quark background samples are initially normalised to their theory predictions.
The $\ttbar$ predicted cross-section is $\sigma_{\ttbar} = 253^{+13}_{-15}$~pb, computed at next-to-next-to-leading-order (NNLO) in QCD, including resummation of next-to-next-to-leading logarithmic (NNLL) soft gluon terms with {\scshape top++2.0}~\cite{Cacciari:2011hy,Beneke:2011mq,Baernreuther:2012ws,Czakon:2012zr,Czakon:2012pz,Czakon:2013goa,Czakon:2011xx}.
The single-top predicted cross-sections are $\sigma_{t}=87.8^{+3.4}_{-1.9}$~pb~\cite{Kidonakis:2011wy} and $\sigma_{\Wt}=22.4\pm 1.5$~pb~\cite{Kidonakis:2010ux}, computed at NLO with NNLL corrections.

The {\scshape Alpgen} v2.14 LO generator~\cite{Mangano:2002ea}, interfaced to {\scshape Pythia} v6.4, is used to generate $W$+jets and $Z$+jets events, with the CTEQ6L1~\cite{Pumplin:2002vw} PDF set.
A parton--jet matching scheme is employed to avoid double-counting of partonic configurations generated by both the matrix-element calculation and the parton shower~\cite{Mangano:2001xp}. 
The samples are generated separately for $W/Z$ with light-quark jets ($W$+light-jets, $Z$+light-jets) and heavy-quark jets ($Wbb$+jets, $Wcc$+jets, $Wc$+jets, $Zbb$+jets, $Zcc$+jets). 
Overlap between the events in samples with heavy quarks generated from the matrix-element calculation and those generated from parton-shower evolution in the samples with light-quark jets is avoided via an algorithm based on the angular separation between the extra heavy quarks, $Q$: if $\Delta R(Q,Q') > 0.4$, the matrix-element prediction is used, otherwise the parton-shower prediction is used.
Diboson events ($WW$, $WZ$, $ZZ$) are generated with {\scshape Alpgen} interfaced to {\scshape Herwig} v6.52~\cite{Corcella:2000bw} for hadronisation and {\scshape Jimmy} v4.31~\cite{Butterworth:1996zw} for the modelling of the underlying event, with the CTEQ6L1 PDF set.

After event generation, most signal and all background samples are passed through the full simulation of the ATLAS detector~\cite{SOFT-2010-01} based on GEANT4~\cite{Agostinelli:2002hh} and reconstructed using the same procedure as for collision data.
A faster simulation~\cite{ATL-PHYS-PUB-2010-013} is used for the signal samples in the dilepton selection and for samples selected to assess systematic uncertainties.
All samples are simulated with additional proton--proton interactions in the same bunch crossing ("pile-up") and reweighted to have the same distribution of the mean number of interactions per bunch-crossing as the data.

\section{Event selection and background estimation}
\label{sec:sel}
\begin{figure}[htbp]
  \begin{center}
\subfigure[]{
\label{fig:Catboosttop}
\includegraphics[scale=0.75]{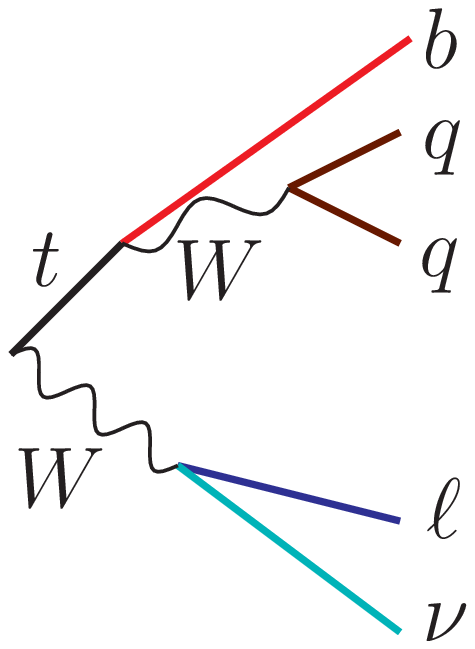}
}
\subfigure[]{
\label{fig:CatboostW2}
\includegraphics[scale=0.75]{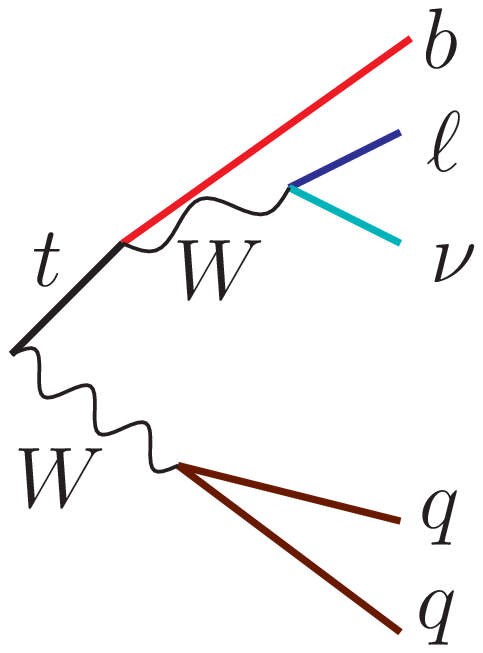} }
\subfigure[]{
\label{fig:Catdilep}
\includegraphics[scale=0.75]{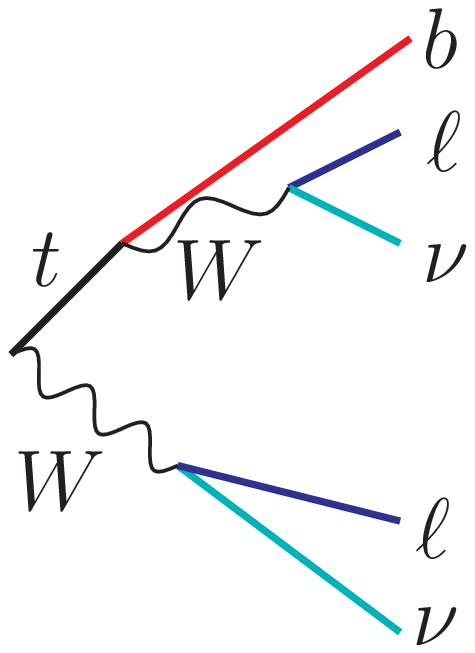} 
}
\caption{Final-state categories:~\subref{fig:Catboosttop} one lepton and a hadronic top-quark decay, \subref{fig:CatboostW2} one lepton and a hadronic $W$-boson decay and \subref{fig:Catdilep} dilepton.
\label{fig:Cat}}
  \end{center}
\end{figure}

Two types of events are selected: those in which the prompt $W$~boson and the one from the top quark both decay leptonically (dilepton final
state), and those in which only one $W$~boson decays leptonically and the other one decays hadronically (lepton+jets final state). Only electrons and muons are considered for the leptonic $W$ decay. 
The events are separated into orthogonal categories based on the decay signature of the two
$W$~bosons, as illustrated in Figure~\ref{fig:Cat}. Full details of the methods used to assign
events to the categories shown in Figure~\ref{fig:Catboosttop} and ~\ref{fig:CatboostW2} are given
below.

\subsection{Object reconstruction}
\label{sec:obj}

Electron candidates are reconstructed from energy clusters in the EM calorimeter and matched to tracks in the inner detector.
Selected electrons must have a transverse energy ${\ET=E_\mathrm{cluster}/\cosh(\eta_\mathrm{track})> \SI{25}{\GEV}}$ and $|\eta_\mathrm{cluster}| < 2.47$, where $E_\mathrm{cluster}$ and
$\eta_\mathrm{cluster}$ indicate the electromagnetic cluster energy and pseudorapidity, respectively, and $\eta_\mathrm{track}$ the track pseudorapidity~\cite{Aad:2014fxa}. 
A veto is placed on electrons in the transition region between the barrel and end-cap calorimeter, $1.37 < |\eta_\mathrm{cluster}| < 1.52$.  
The electron track is required to originate less than \SI{2}{mm} along the $z$-axis (longitudinal impact parameter) from the selected event primary vertex.
The three main sources of background for high-\ET\ isolated electrons are hadrons misidentified as electrons, photon conversions and electrons originating from secondary vertices in decays of heavy-flavour hadrons (non-prompt electrons).
In order to suppress the background from these sources, it is required that there is little calorimeter activity in the region surrounding the electron candidate. 
Two isolation variables are defined for this purpose: the energy deposited around the electron candidate in the calorimeter
with a cone size of $\Delta R = 0.2$, and the sum of the transverse momenta of all tracks in a cone of size $\Delta R =0.3$ around the electron candidate. 
Efficiencies and purities of the identification and isolation requirements are corrected with appropriate scale factors to match the simulation to the data. 
They depend on the $\eta$ and the \ET\ of the electron. 

Muon candidates are reconstructed by matching muon spectrometer hits to inner-detector tracks, using the complete track
information from both detectors and accounting for scattering and energy loss in the ATLAS detector material. 
Selected muons have a transverse momentum greater than \SI{25}{\GEV} and a pseudorapidity of $|\eta|<2.5$. 
The muon track longitudinal impact parameter with respect to the primary vertex is required to be less than \SI{2}{mm}.
Isolation criteria are applied in order to reduce background contamination from events in which a muon is produced at a secondary vertex in the decay of a heavy-flavour quark (non-prompt muon).
The ratio of the summed \pT\ of tracks within a cone of variable size $\Delta R = \SI{10}{\GEV} / \pt^{\mu}$ to the \pT\ of the muon is required to be less than 0.05~\cite{Rehermann:2010vq}.
Efficiencies and purities of the identification and isolation requirements are corrected with appropriate scale factors to match the simulation to the data; they depend on $\eta$ and $\phi$ of the muon. 

Jets are reconstructed using the anti-$k_{t}$ algorithm~\cite{Cacciari:2008gp}, applied to clusters of calorimeter cells that are topologically connected and
calibrated to the hadronic energy scale~\cite{PERF-2011-03} using a local calibration scheme~\cite{Issever:2004qh}. Jet energies are calibrated using energy- and $\eta$-dependent correction factors derived from simulation, and with residual corrections from in-situ measurements~\cite{Aad:2014bia}. Events with jets built from noisy calorimeter cells or non-collision background processes are removed~\cite{Aad:2013gja}.

In this paper jets reconstructed with two different radius parameters are used. Small-$R$ jets (also denoted simply by ``jets'' or $j_{4}$) have a radius parameter of 0.4 and are required to have $\pT > \SI{25}{\GEV}$ and $|\eta| <4.5$.
Small-$R$ jets from additional simultaneous $pp$ interactions are rejected by an additional requirement: the ratio of the scalar sum of the $\pT$ of tracks associated with the jet and the primary vertex to the scalar sum of the $\pT$ of all tracks associated with the jet must be at least 0.5 for jets with $\pt<\SI{50}{\GEV}$ and $|\eta|<2.4$.
Large-$R$ jets ($j_{10}$) have a radius parameter of 1.0~\cite{Aad:2013gja} and are subject to a trimming procedure~\cite{Krohn:2009th} to minimise the impact of energy depositions from pile-up interactions. 
The trimming algorithm reconstructs constituent jets with the $k_t$ algorithm~\cite{Catani:1993hr,Ellis:1993tq,Cacciari:2011ma} with a radius parameter of 0.3 from the clusters belonging to the original large-$R$ jet. 
Constituent jets contributing less than 5\% of the large-$R$ jet $\pt$ are removed. The remaining energy depositions are used to calculate the kinematic and
substructure properties of the large-$R$ jet. The masses of the large-$R$ jets ($m_{j_{10}}$) are calibrated to
their particle-level values~\cite{Aad:2014bia,STDM-2011-19,ATLAS-CONF-2012-065}. In the analysis, large-$R$ jets are required to have $\pT > \SI{200}{\GEV}$ and $|\eta| <2.0$.

Since the reconstruction of jets and electrons is partially based on the same energy clusters, the closest small-$R$ jet overlapping with electron candidates within a cone of size  $\Delta R = 0.2$ is removed.
Since electrons from prompt $W$-boson decays should be isolated from jet activity, electrons that are still within less than $\Delta R = 0.4$ of a jet are removed.
Similarly, prompt muons should be isolated from jet activity, so that beside the muon-isolation criteria above, selected muons must not overlap with a reconstructed jet within a radius of size $\Delta R = 0.4$.

Small-$R$ jets containing $b$-quarks ($b$-quark jets) from the top-quark decay are identified using a combination of multivariate algorithms ($b$-tagging)~\cite{ATLAS-CONF-2014-046}. 
These algorithms exploit the long lifetime of $B$~hadrons and the properties of the resulting displaced and reconstructed secondary vertex.
The working point of the algorithm is chosen such that $b$-quark jets in simulated $\ttbar$ events are tagged with an average efficiency of $70\%$ and a rejection
factor for light-flavour jets of $\sim 100$.

The missing transverse momentum vector, $\vmet$, is calculated as the negative vector sum over all the calorimeter cells in the event, and is further refined by applying object-level corrections for the contributions which arise from identified photons, leptons and jets~\cite{Aad:2012re}. 
In the analysis the magnitude, $\met$, of the missing transverse momentum vector is used.

\subsection{Single-lepton event selection and background estimation}
\label{sec:sellj}

\begin{figure}[!h!tbp]
\begin{center}\
\subfigure[]{
\label{F:leading_fatjet_mass_oneL_bprime_extra}
\includegraphics[width=0.47\textwidth]{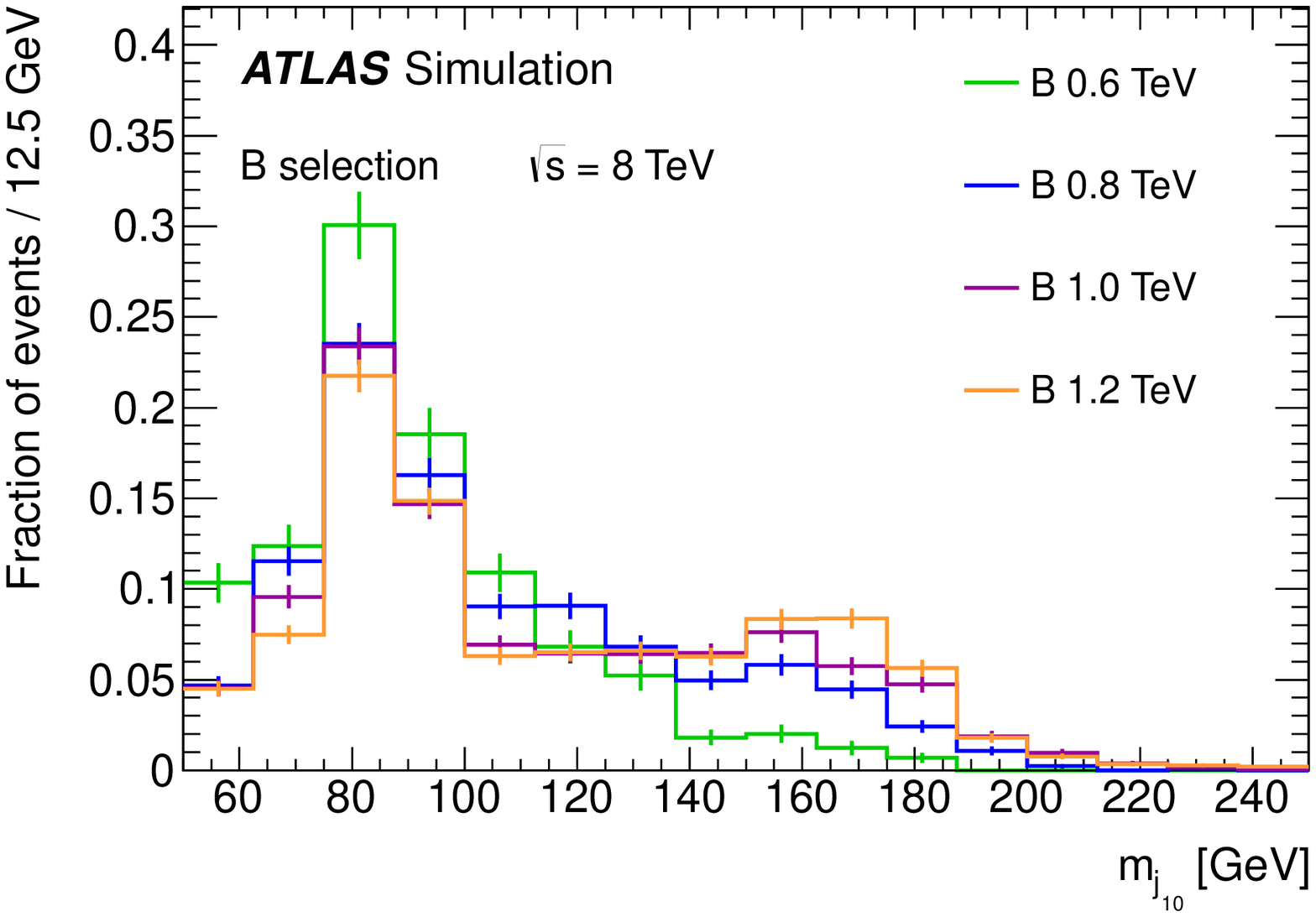}} 
\subfigure[]{
\label{F:leading_fatjet_mass_oneL_bstar_extra}
\includegraphics[width=0.47\textwidth]{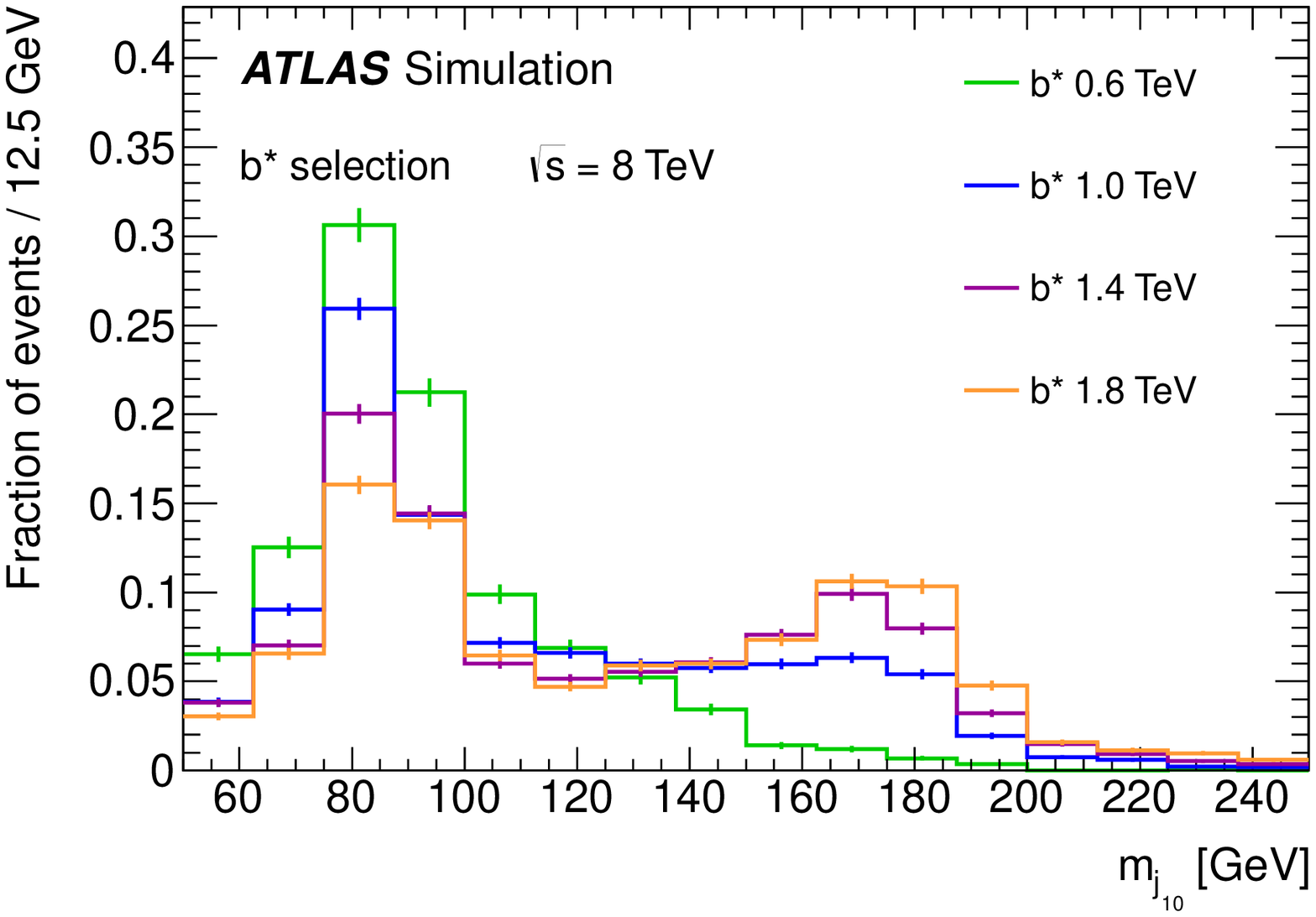} }

\caption{Comparison of the leading large-$R$ jet mass for various simulated \subref{F:leading_fatjet_mass_oneL_bprime_extra} \Bprime\ and
\subref{F:leading_fatjet_mass_oneL_bstar_extra} \bstar\ signal masses before applying any requirement on angular distances which later define the signal regions.
The distributions are normalised to unity.
\label{fig:leading_fatjet_mass_oneL}}
\end{center}
\end{figure}

In the single-lepton final state, the selected events have exactly one isolated lepton (electron or muon) and exactly two or three small-$R$ central jets ($|\eta| < 2.5$), exactly one of
which is $b$-tagged. Events with two jets are included in order to select events where the two jets from the $W$~boson are merged at larger boost. Events are also required to have at least one large-$R$ jet with $\pt^{j_{10}} > \SI{200}{\GEV}$ and $m_{j_{10}} > \SI{50}{\GEV}$.
If several massive large-$R$ jets are found, the one with the highest $\pt$ is considered. Requirements on the missing transverse momentum and the transverse mass of the lepton--$\met$ system\footnote{
The transverse mass is defined as $\mtw = \sqrt{2\pt(\ell)\met[1-\cos\Delta\phi(\pt(\ell),\met)]}$, where $\pt$ denotes the magnitude of the lepton transverse momentum, and $\Delta\phi(\pt(\ell),\met)$ the azimuthal difference between the missing transverse momentum and the lepton direction.}
of ${\met> \SI{20}{\GEV}}$ and $\mtw+\met > \SI{60}{\GEV}$, respectively, reduce the fraction of selected events originating from non-prompt
or misidentified leptons. A requirement on the azimuthal angle between the large-$R$ jet and the
lepton, $\Delta \phi(\ell, j_{10}) > 1.5$, increases the signal-to-background ratio, because in signal events large-$R$ jets from boosted hadronic top-quark ($W$-boson) decays recoil against the leptonic decay of
the other $W$~boson (top quark) as shown in Figure~\ref{fig:Catboosttop} and ~\ref{fig:CatboostW2}.
Figure~\ref{fig:leading_fatjet_mass_oneL} shows the mass distribution of the leading large-$R$ jet at this selection level for various
simulated \Bprime\ (left) and \bstar\ (right) signal masses studied in this paper.

Angular correlations are used to select single-lepton events and to categorise them in the different signal regions: 
if the smallest distance between the lepton and any small-$R$ jet satisfies $\Delta R (\ell, j_{4}) > 1.5$, and if the largest
distance between the leading (highest \pt) large-$R$ jet and any small-$R$ jet satisfies $\Delta R (j_{4}, j_{10}) < 2.0$, events are assigned to the 1L hadronic top region
(\textit{1L hadT}, see Figure~\ref{fig:Catboosttop}). However, if the smallest distance between the lepton and any small-$R$ jet is $\Delta R(\ell, j_{4}) < 1.5$
and the largest distance between the leading large-$R$ jet and any small-$R$ jet is $\Delta R (j_{4}, j_{10}) > 2.0$, the event is
assigned to the 1L hadronic $W$ category (\textit{1L hadW}, see Figure~\ref{fig:CatboostW2}).
If an event cannot be assigned to any of the two categories, it is rejected.

For the \Bprime\ and \bstar\ signal regions the same selection cuts are applied, except that for the \Bprime\ signal regions at least one additional forward small-$R$ jet ($2.5 <|\eta| < 4.5$)
is required. The selection with (without) forward-jet requirement is referred to as \Bprime\ (\bstar) selection.

\begin{figure}[!h!tbp]
\begin{center}
\subfigure[]{
\label{F:SR_lep_pt_only_hadTop}
\includegraphics[width=0.48\textwidth]{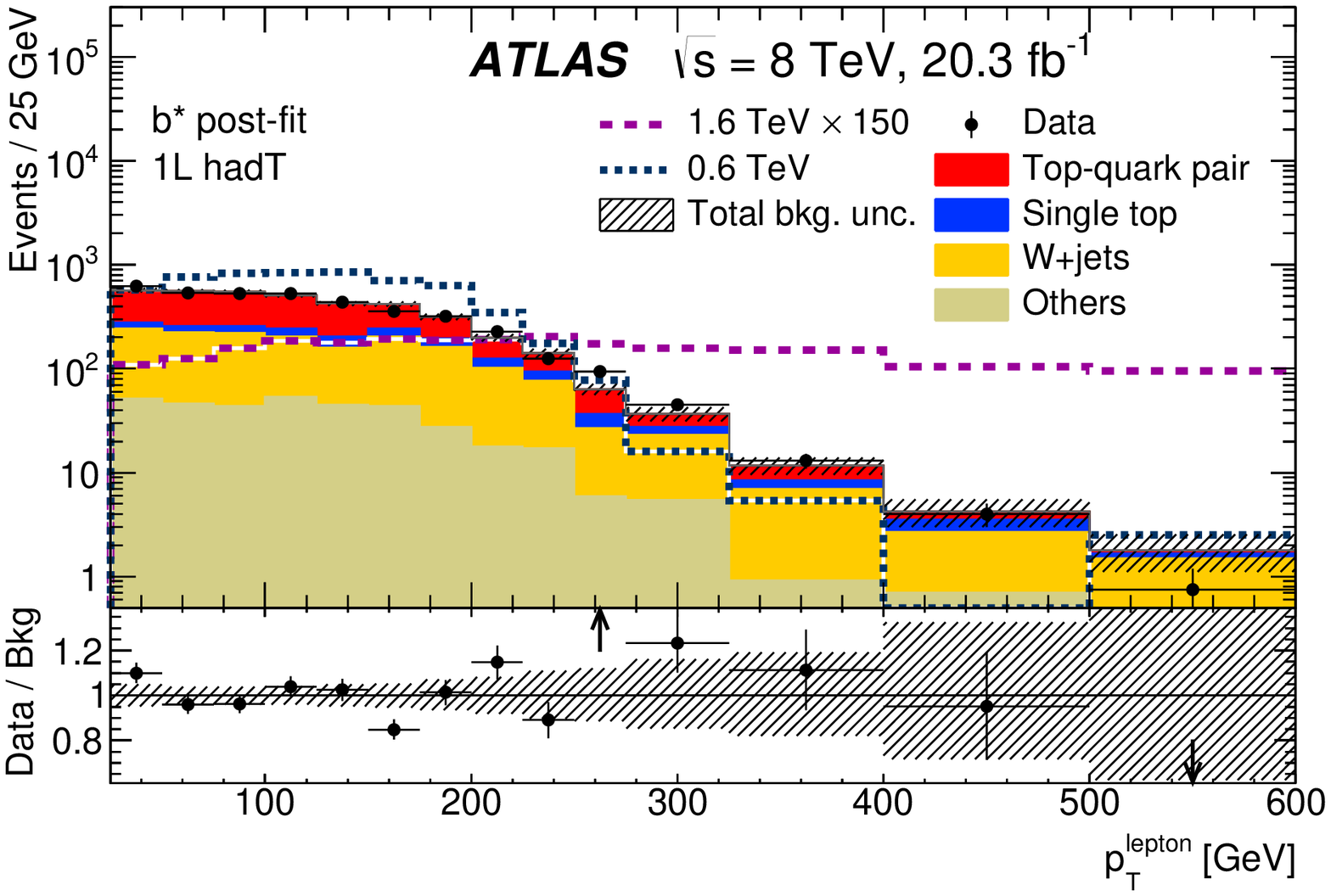} }
\subfigure[]{
\label{F:SR_lep_pt_only_hadW}
\includegraphics[width=0.48\textwidth]{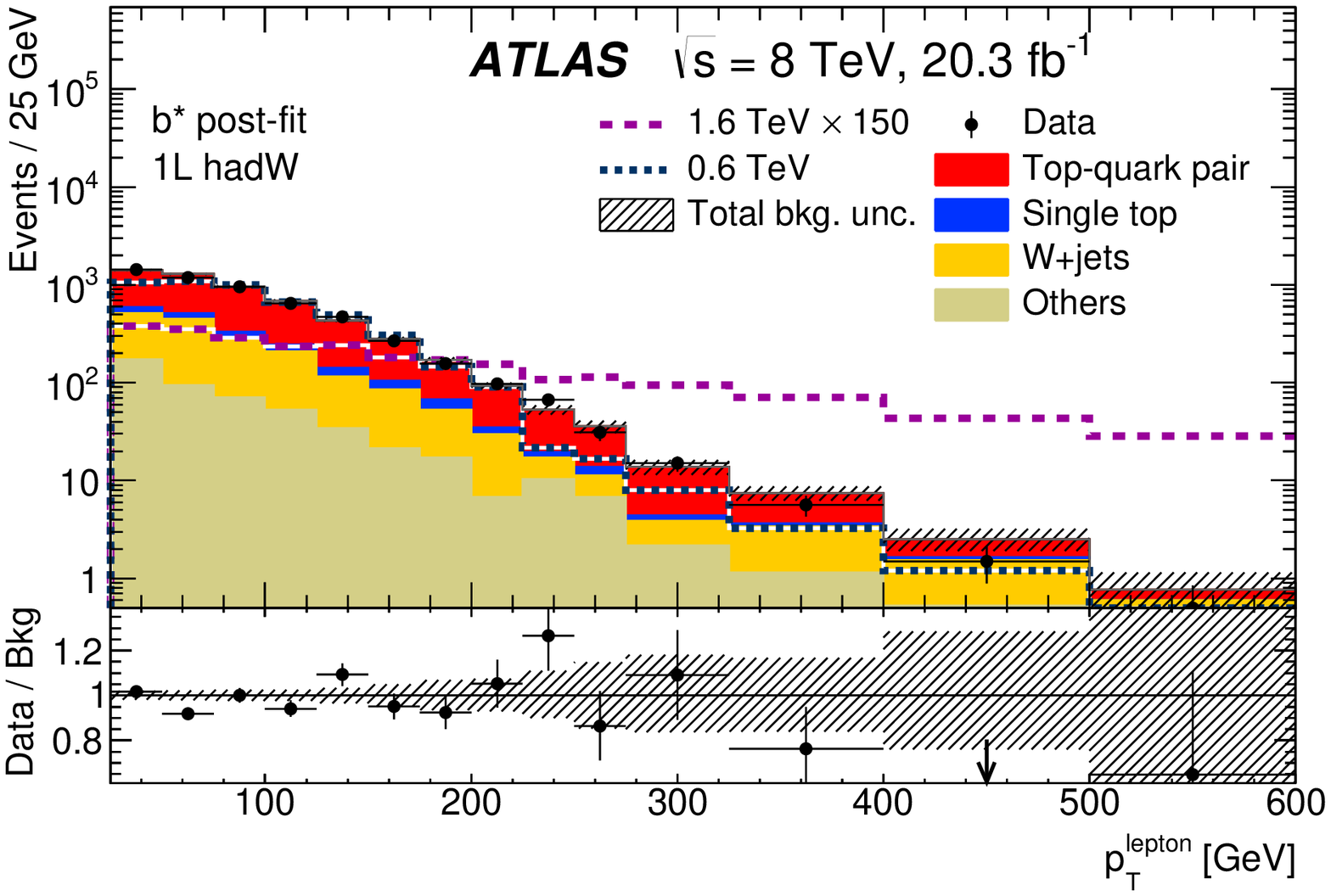} }
\caption{Distributions of the lepton transverse momentum in the \bstar\ selection of the single-lepton analysis for the data, 
SM background processes and two \bstar\ signal masses in the \subref{F:SR_lep_pt_only_hadTop} \bstar\ \textit{1L hadT} and \subref{F:SR_lep_pt_only_hadW}
\bstar\ \textit{1L hadW} signal regions. 
The signal distributions are scaled to the theory cross-sections multiplied by the indicated factor for better visibility and the background processes are normalised to their post-fit normalisation.
The uncertainty band includes statistical and all systematic uncertainties. The last bin includes the overflow.
\label{fig:lplus_jets_SR_lep_pt}}

\end{center}
\end{figure}

\begin{figure}[!h!tbp]
\begin{center}

\subfigure[]{
\label{F:SR_first_FWjet_pt_only_FWjet_hadTop}
\includegraphics[width=0.48\textwidth]{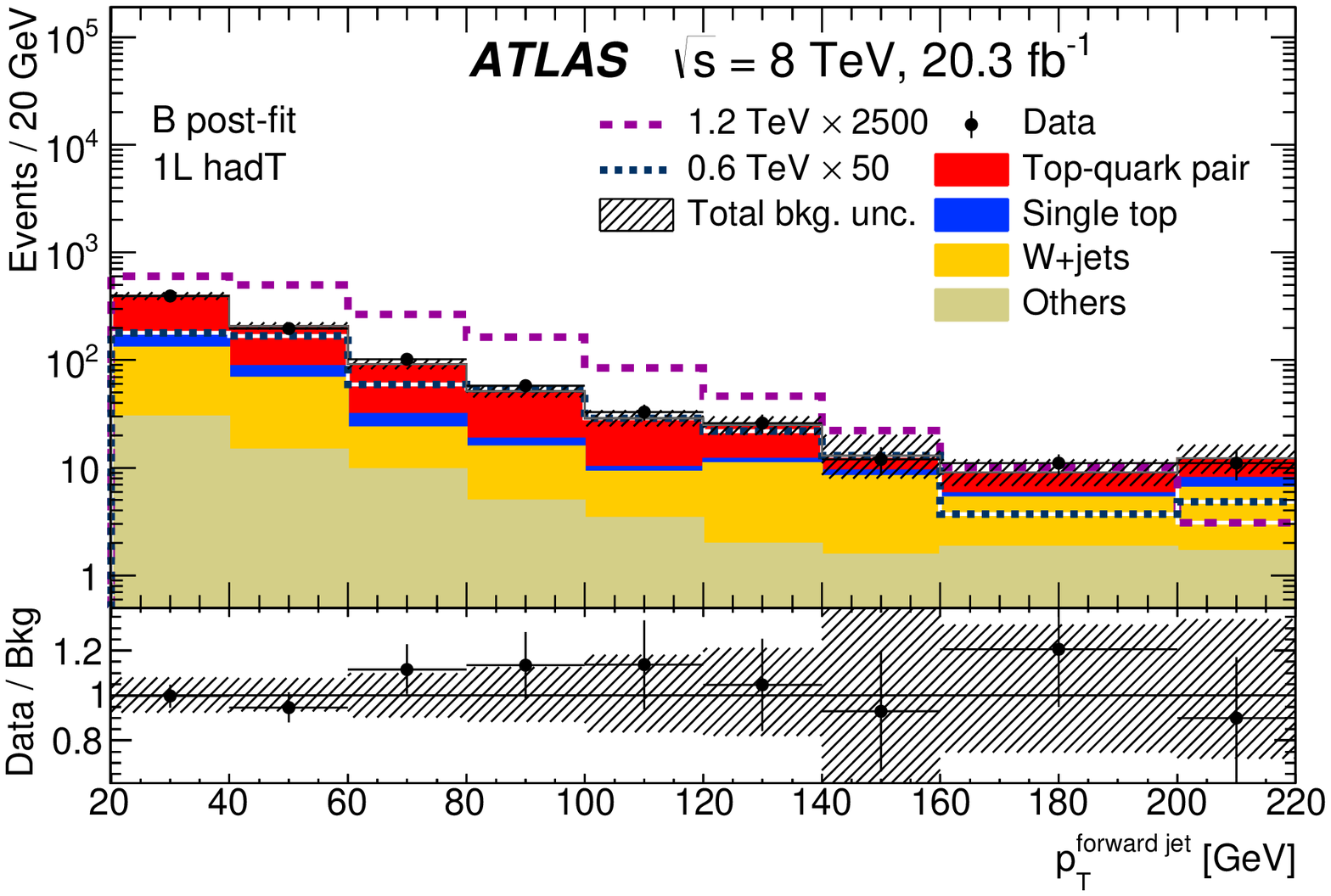} }
\subfigure[]{
\label{F:SR_first_FWjet_pt_only_FWjet_hadW}
\includegraphics[width=0.48\textwidth]{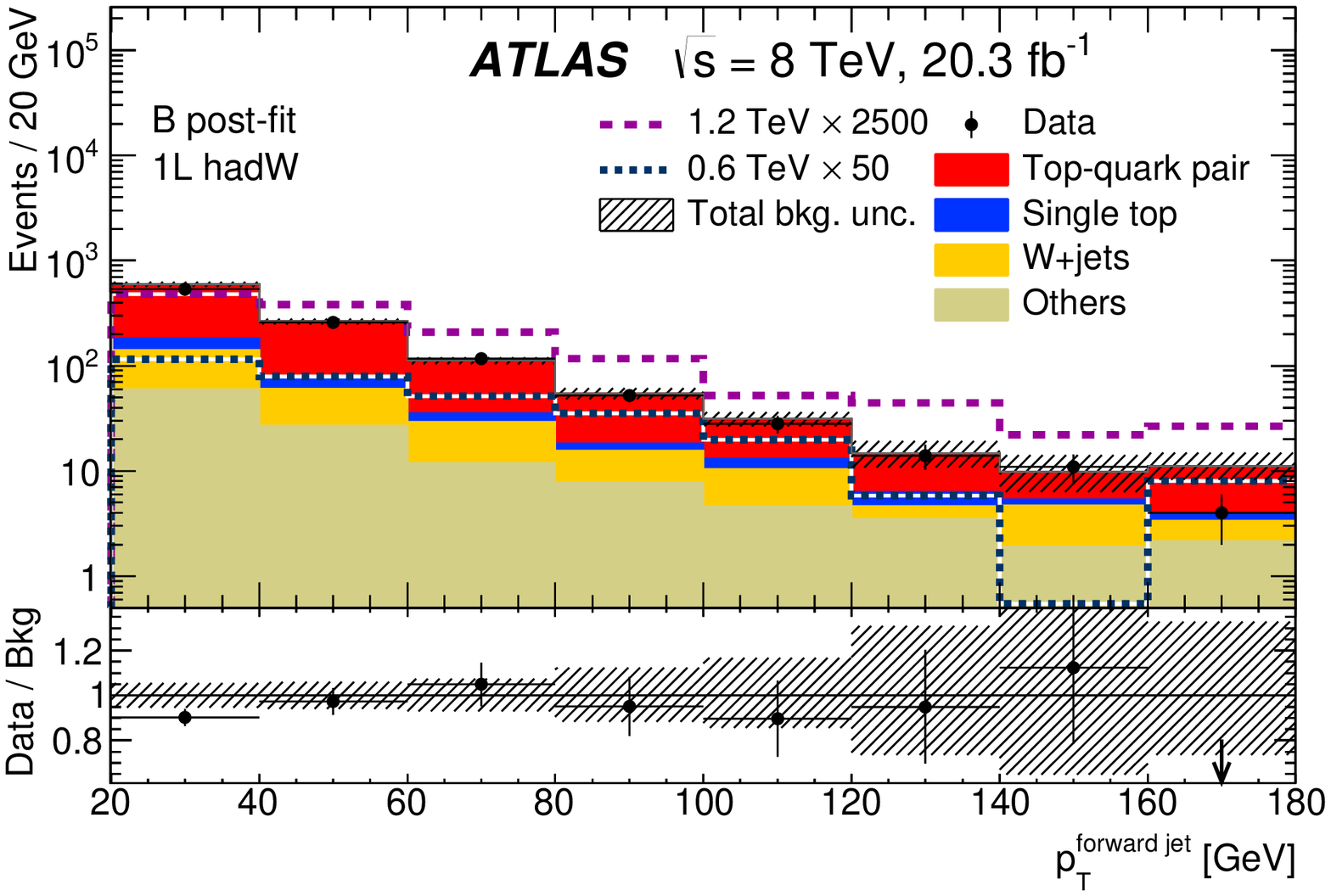} }

\caption{Distributions of the leading forward-jet transverse momentum in the \Bprime\ selection of the single-lepton analysis
for the data events, SM background processes and two \Bprime\ signal masses in the \subref{F:SR_first_FWjet_pt_only_FWjet_hadTop} \Bprime\ \textit{1L hadT} and \subref{F:SR_first_FWjet_pt_only_FWjet_hadW} \Bprime\ \textit{1L hadW} signal region. 
The signal distributions are scaled to the theory cross-sections multiplied by the indicated factor for better visibility and the background processes are normalised to their post-fit normalisation.
The uncertainty band includes statistical and all systematic uncertainties. The last bin includes the overflow.
\label{fig:lplus_jets_SR_first_FWjet_pt}}
\end{center}
\end{figure}

Figure~\ref{fig:lplus_jets_SR_lep_pt} presents the distributions of the lepton transverse momentum for the \bstar\ selection of the
single-lepton analysis for the predicted SM background processes, for data events and two \bstar\ signal masses studied in this paper
in the \textit{1L hadT} (left) and \textit{1L hadW} (right) regions. The transverse momentum distribution of the leading (highest \pt) forward jet after applying all selection criteria in the two \Bprime\ signal
regions is shown in Figure~\ref{fig:lplus_jets_SR_first_FWjet_pt}. The background processes are normalised to the post-fit event yields (see Section~\ref{sec:res}) and the
data agrees with the SM expectation within the statistical and systematic uncertainties band in both figures.

The main background processes with a single-lepton signature arise from top-quark pair and $W$+jets production, with smaller contributions from single top-quark production, multijet, $Z$+jets and diboson events. 
The normalisation of $W$+jets processes as well as the $\ttbar$ and single-top $\Wt$-processes is improved by constraining the predicted yields to match the data in  appropriate control regions enriched in $W$ and $\ttbar$ events, respectively.
Two control regions (CRs) for each signal region are defined for this purpose. The selection for the $W$-enriched region is
the same as for the signal region, except that each event is required to have exactly zero $b$-tagged jets. The selection for the $\ttbar$ control regions is
the same as for the signal regions, except that only events with two or more $b$-tagged jets are selected.
The multijet processes are estimated with a data-driven matrix method described in reference~\cite{ATLAS-CONF-2014-058}.
The remaining background processes, such as single top-quark production, $Z$+jets and diboson processes are estimated using simulations. 
The contamination of the $\ttbar$ control regions by \bstar\ and \Bprime\ signals with masses of $m{\geq \SI{1000}{\GEV}}$ is at most 16\% and 0.9\%,
respectively, and in the $W$-enriched control regions at most 2.2\% and 0.1\%, respectively.
\newline

\begin{figure}[htb!!!]
\begin{center}
\subfigure[]{
\label{F:m_tot_only_hadTop}
\includegraphics[width=0.48\textwidth]{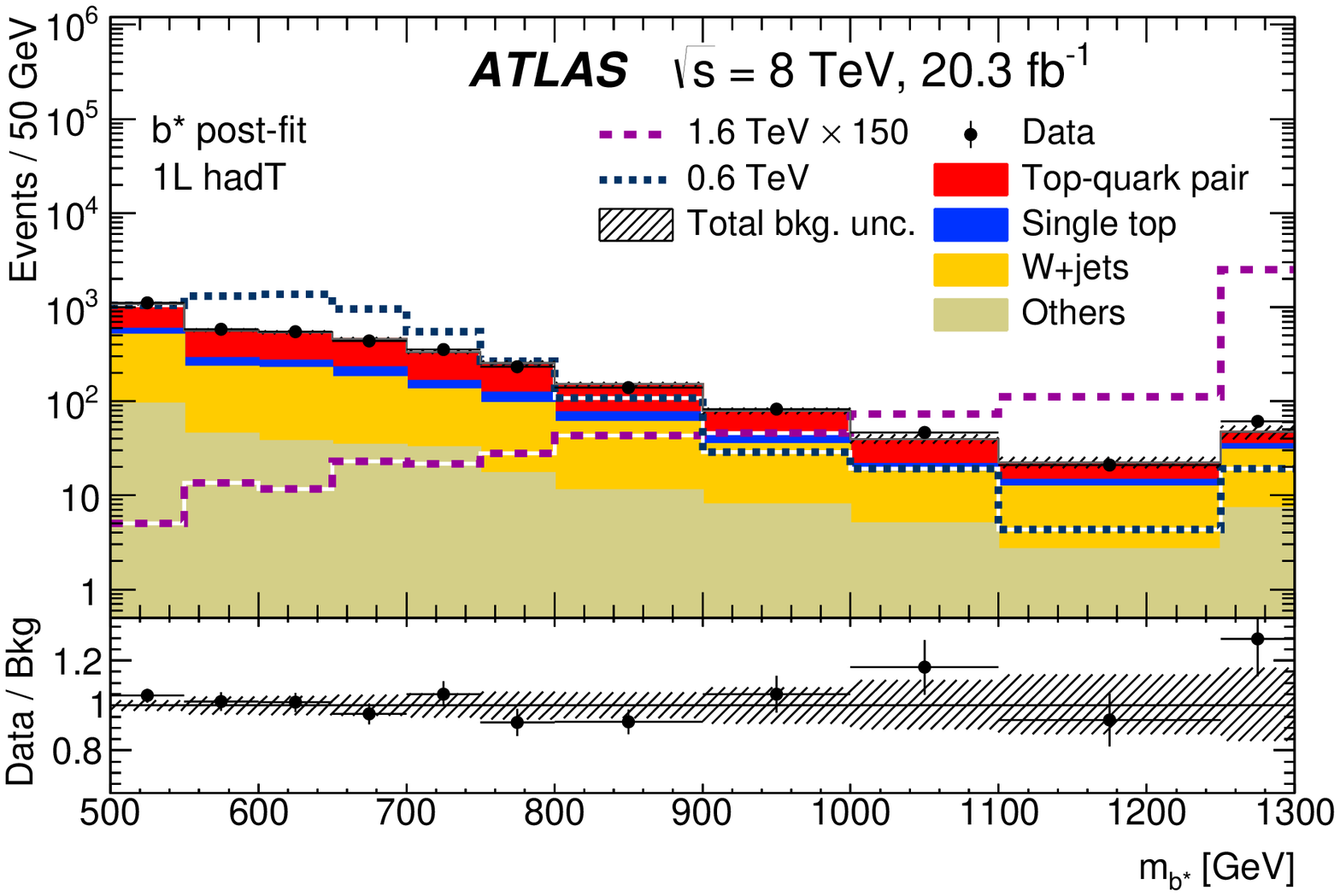} }
\subfigure[]{
\label{F:m_tot_only_hadW}
\includegraphics[width=0.48\textwidth]{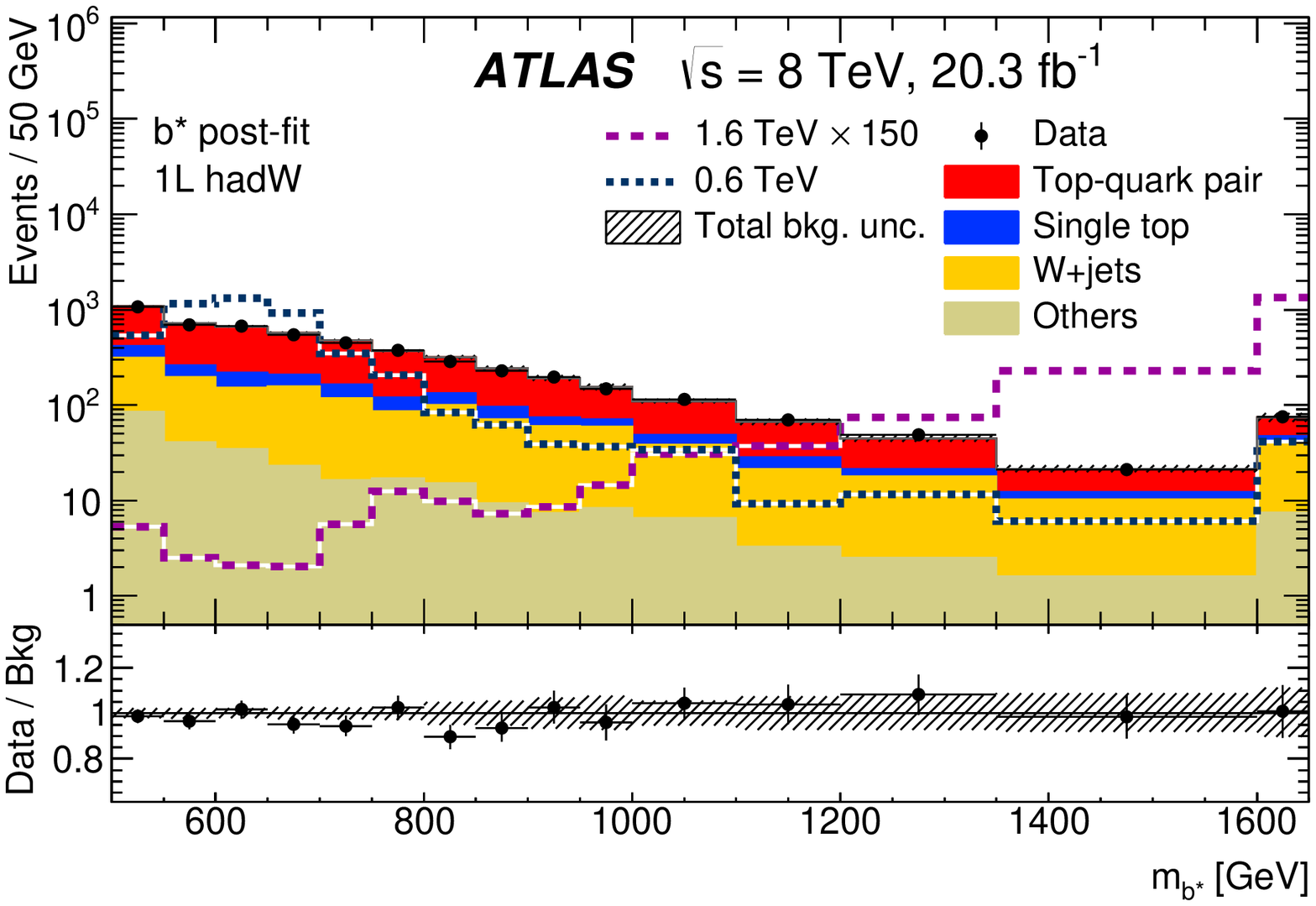} }
\subfigure[]{
\label{F:m_tot_only_FWjet_hadTop}
\includegraphics[width=0.48\textwidth]{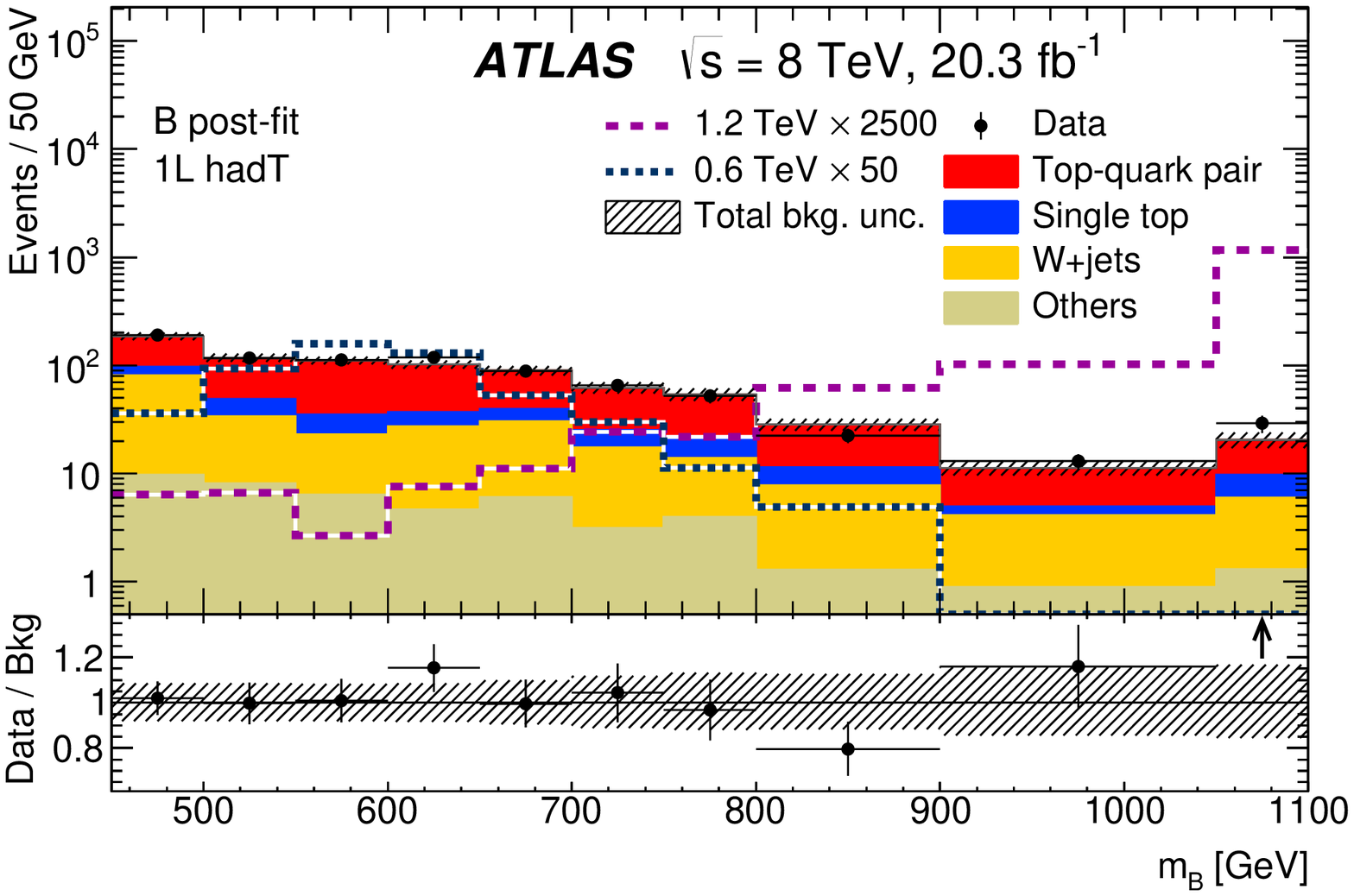} }
\subfigure[]{
\label{F:m_tot_only_FWjet_hadW}
\includegraphics[width=0.48\textwidth]{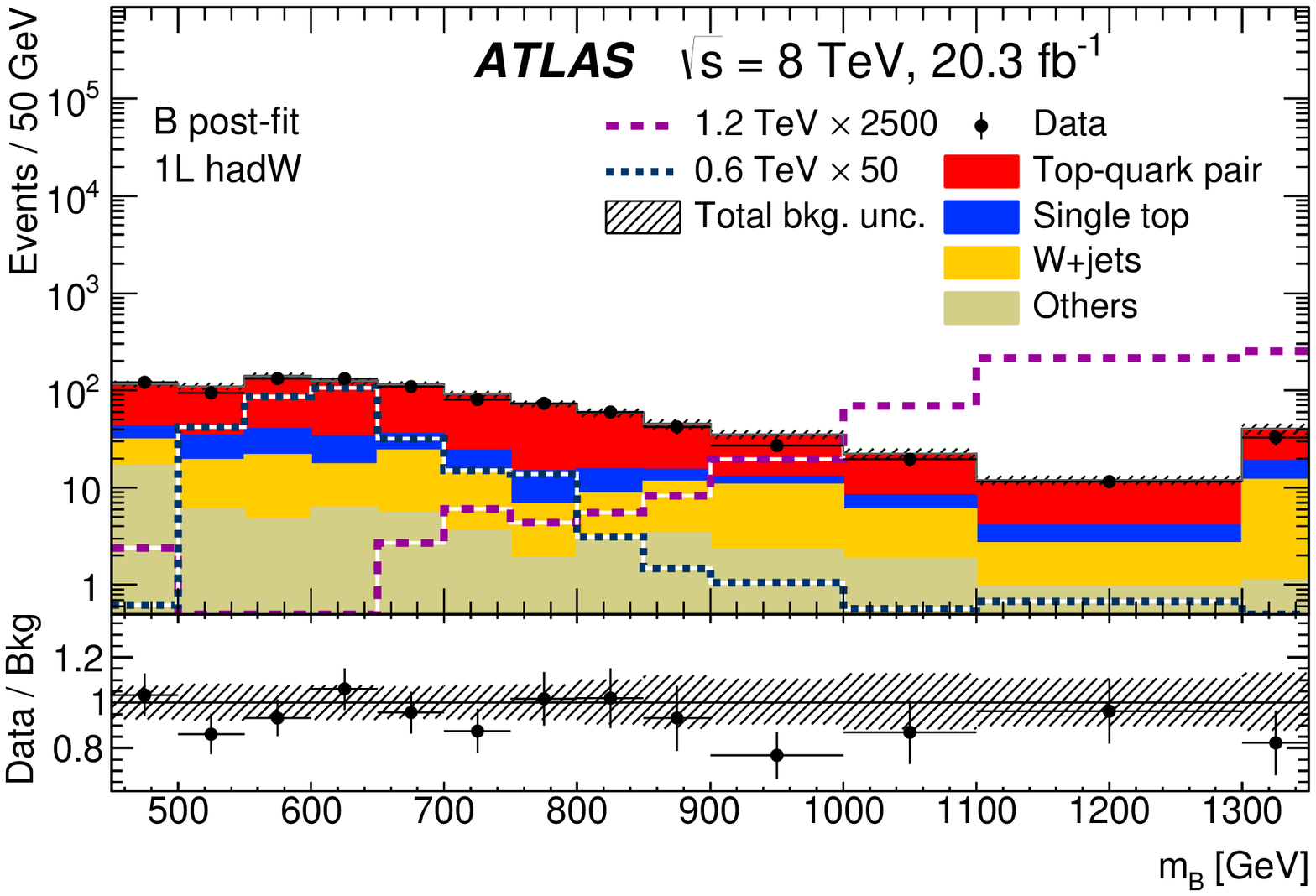} }
\caption{Distributions of the reconstructed mass in the \bstar\ (top row) and \Bprime\ (bottom row) selection of the single-lepton analysis for the SM background processes, data and two \bstar\ (top row) or \Bprime\ (bottom row) signal masses
in the \subref{F:m_tot_only_hadTop} \bstar\ \textit{1L hadT}, \subref{F:m_tot_only_hadW} \bstar\ \textit{1L hadW}, \subref{F:m_tot_only_FWjet_hadTop} \Bprime\ \textit{1L hadT} and \subref{F:m_tot_only_FWjet_hadW} \Bprime\ \textit{1L hadW} signal
regions. The signal distributions are scaled to the theory cross-sections multiplied by the indicated factor for better visibility and the background processes are normalised to their post-fit normalisation.
The uncertainty band includes statistical and all systematic uncertainties. The first and last bin include the underflow and overflow, respectively.
The binning of the figures is the same as used in the binned likelihood fit function. 
\label{fig:lplus_jets_m_tot_SR_results}}
\end{center}
\end{figure}

\begin{figure}[!h!tbp]
\begin{center}

\subfigure[]{
\label{F:SR_lead_fatjet_m_only_hadTop}
\includegraphics[width=0.48\textwidth]{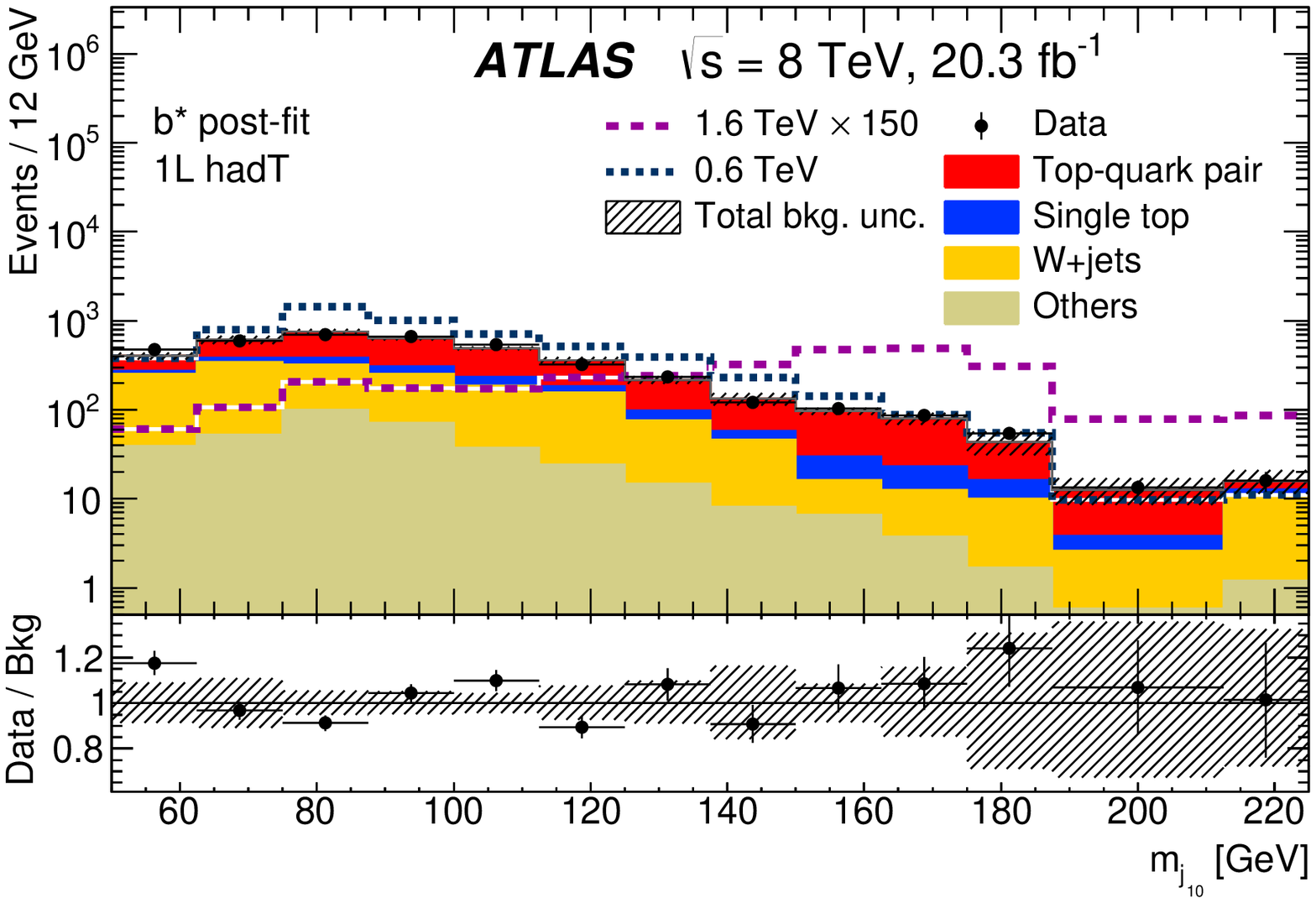} }
\subfigure[]{
 \label{F:SR_lead_fatjet_m_only_hadW}
\includegraphics[width=0.48\textwidth]{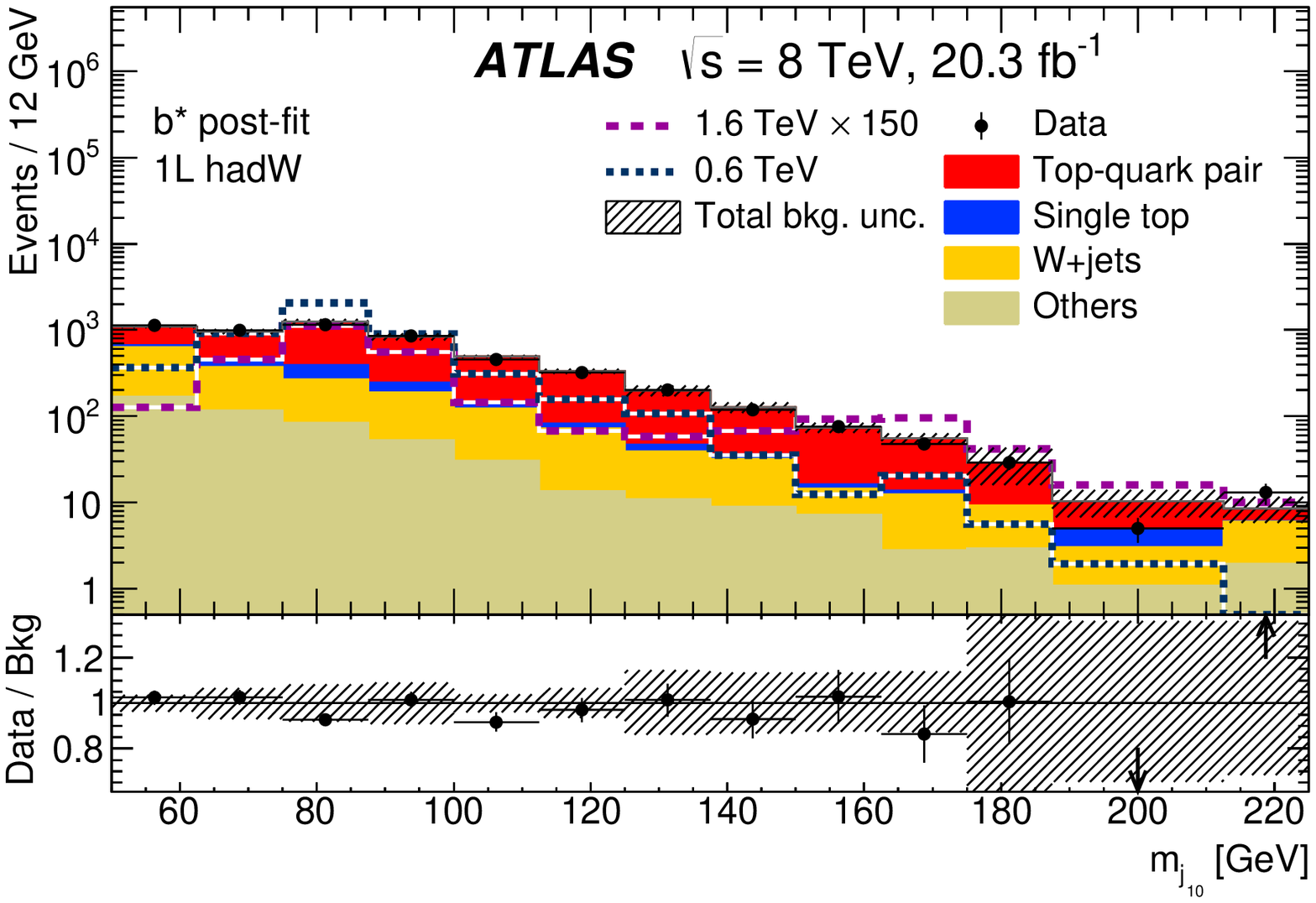} }
\subfigure[]{
\label{F:SR_lead_fatjet_pt_only_FWjet_hadTop}
\includegraphics[width=0.48\textwidth]{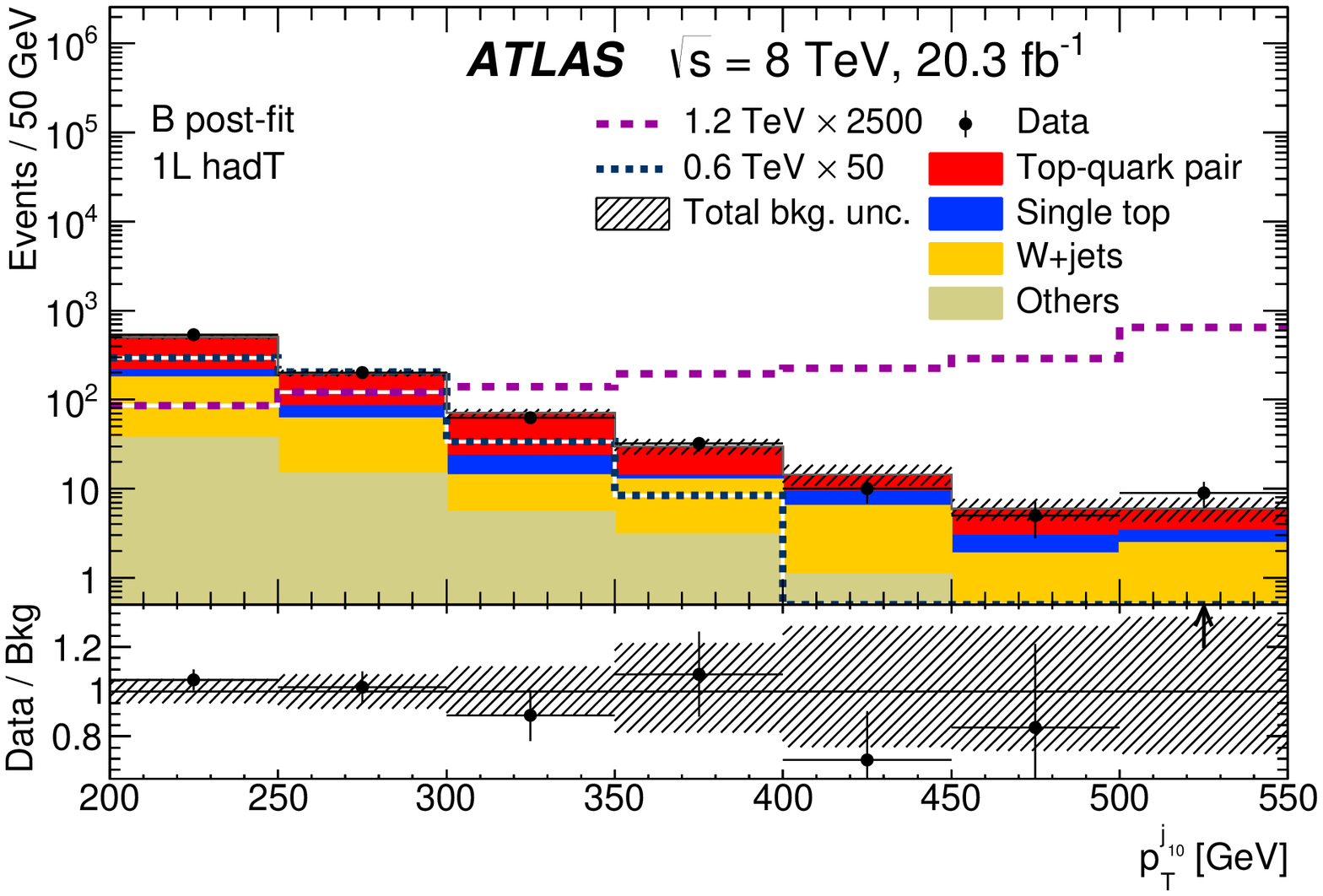} }
\subfigure[]{
\label{F:SR_lead_fatjet_pt_only_FWjet_hadW}
\includegraphics[width=0.48\textwidth]{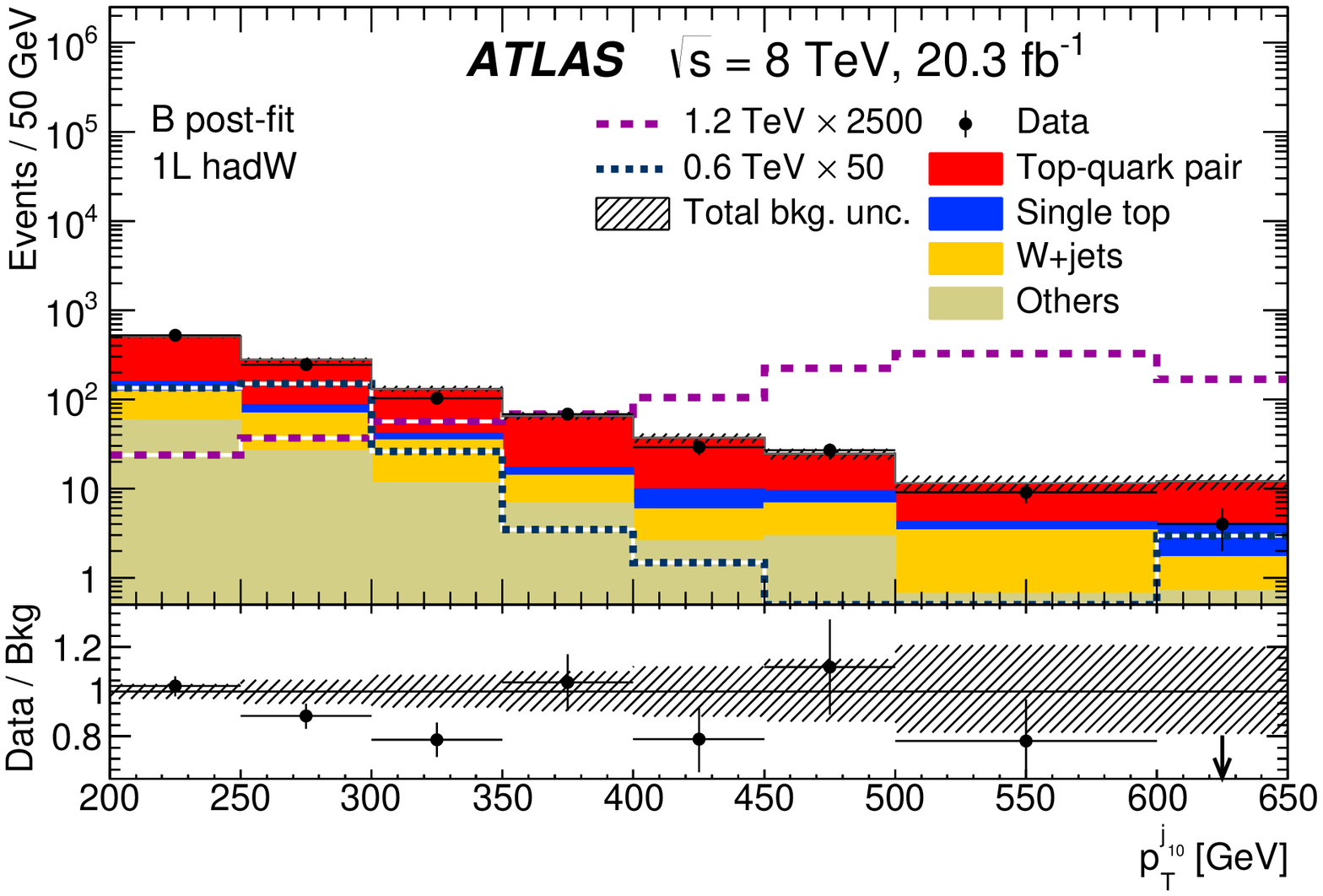} }

\caption{Distributions of the leading large-$R$ jet mass in the \bstar\ selection (top row) and distributions of the leading large-$R$
transverse momentum in the \Bprime\ selection (bottom row) of the single-lepton analysis 
for the SM background processes, data and two \bstar\ (top row) or \Bprime\ (bottom row) signal masses
in the \subref{F:SR_lead_fatjet_m_only_hadTop},\subref{F:SR_lead_fatjet_pt_only_FWjet_hadTop} \textit{1L hadT} and
\subref{F:SR_lead_fatjet_m_only_hadW},\subref{F:SR_lead_fatjet_pt_only_FWjet_hadW} \textit{1L hadW} signal regions. 
The signal distributions are scaled to the theory cross-sections multiplied by the indicated factor for better visibility and the background processes are normalised to their post-fit normalisation.
The uncertainty band includes statistical and all systematic uncertainties. The last bin includes the overflow.
\label{fig:lplus_jets_SR_lead_fatjet_m_pt}}

\end{center}
\end{figure}

The invariant mass calculated from the four-momenta of the lepton, all (two or three) central small-$R$ jets and the missing transverse
momentum is used to discriminate the signal from the background processes in the statistical analysis. Since the longi\-tudinal component of the neutrino momentum is not reconstructed, it is set to zero. Figure~\ref{fig:lplus_jets_m_tot_SR_results} shows the reconstructed mass distribution for
the $\bstar$ selection (top row) and for the $\Bprime$ selection (bottom row) in the \textit{1L hadT} (left side) and the \textit{1L hadW} (right side)
signal region. Further kinematic distributions, such as the leading large-$R$ jet mass in the \bstar\ signal regions and the leading large-$R$ jet
transverse momentum in the \Bprime\ signal regions, are presented in Figure~\ref{fig:lplus_jets_SR_lead_fatjet_m_pt}.
All SM background processes in Figure~\ref{fig:lplus_jets_SR_lep_pt}--\ref{fig:lplus_jets_SR_lead_fatjet_m_pt} are normalised to the post-fit normalisation that is discussed in Section~\ref{sec:res}.
The figures show good agreement between the data and the SM background processes.
The total number of data events and the event yields after fitting the background-only hypotheses to data (as explained in Section~\ref{sec:res}), together with their systematic uncertainties in the \Bprime\ and \bstar\ signal
regions, are listed in Tables~\ref{TABLE-EVENT-YIELD-Bprime} and~\ref{TABLE-EVENT-YIELD-bstar}.

The post-fit normalisation is consistent with the pre-fit expectation within
uncertainties: the normalisation of the top-quark background differs by 5--10\%,
the normalisation of $W$+jets processes in the SRs is consistent with $1.0$,
except for the \textit{1L hadW} B signal region, where it decreases to $0.65 \pm 0.11$.

\begin{figure}[!h!tbp]
\begin{center}
\subfigure[]{
\label{F:WjetsCR_m_tot_only_hadTop}
\includegraphics[width=0.48\textwidth]{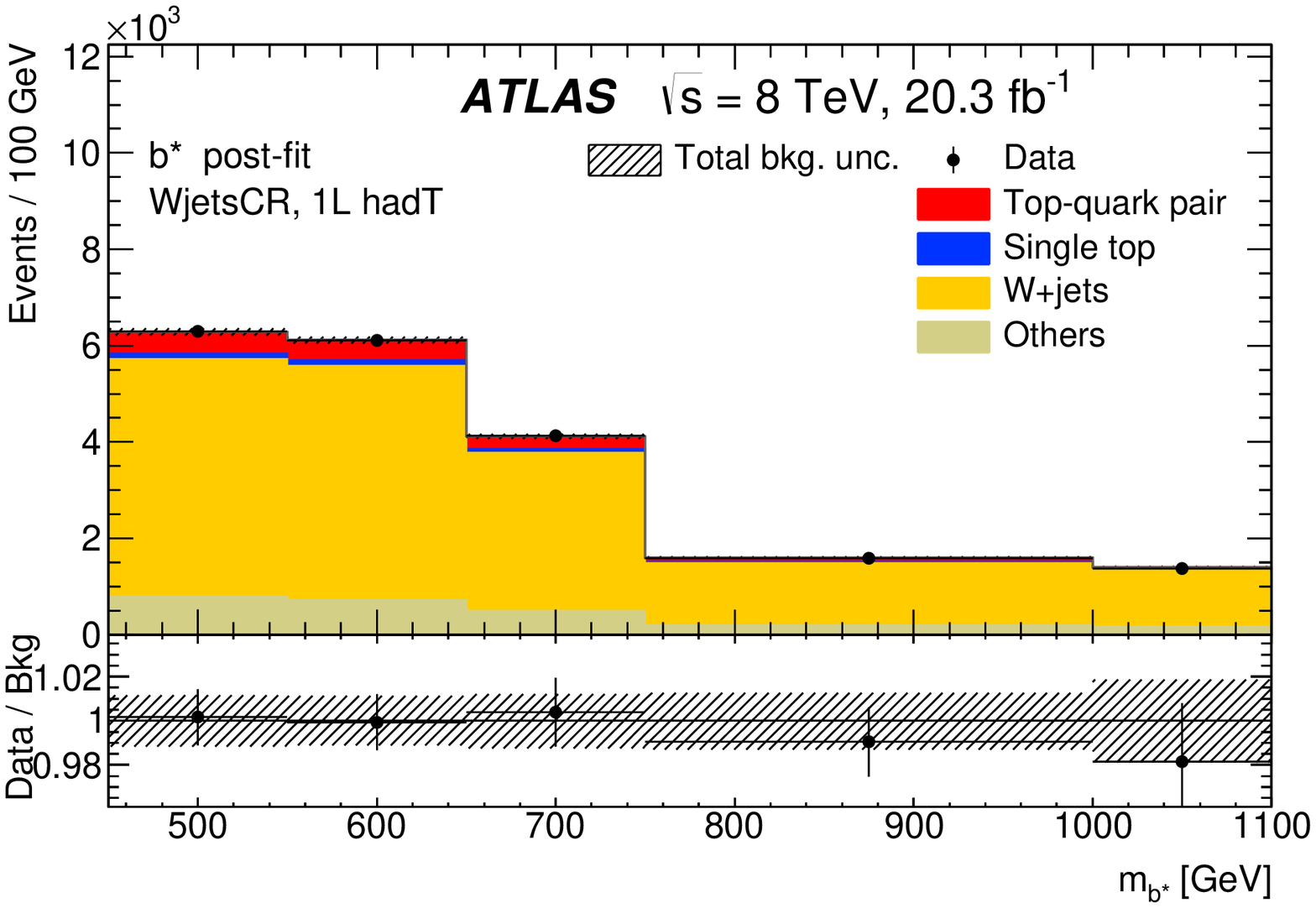} }
\subfigure[]{
\label{F:WjetsCR_m_tot_only_hadW}
\includegraphics[width=0.48\textwidth]{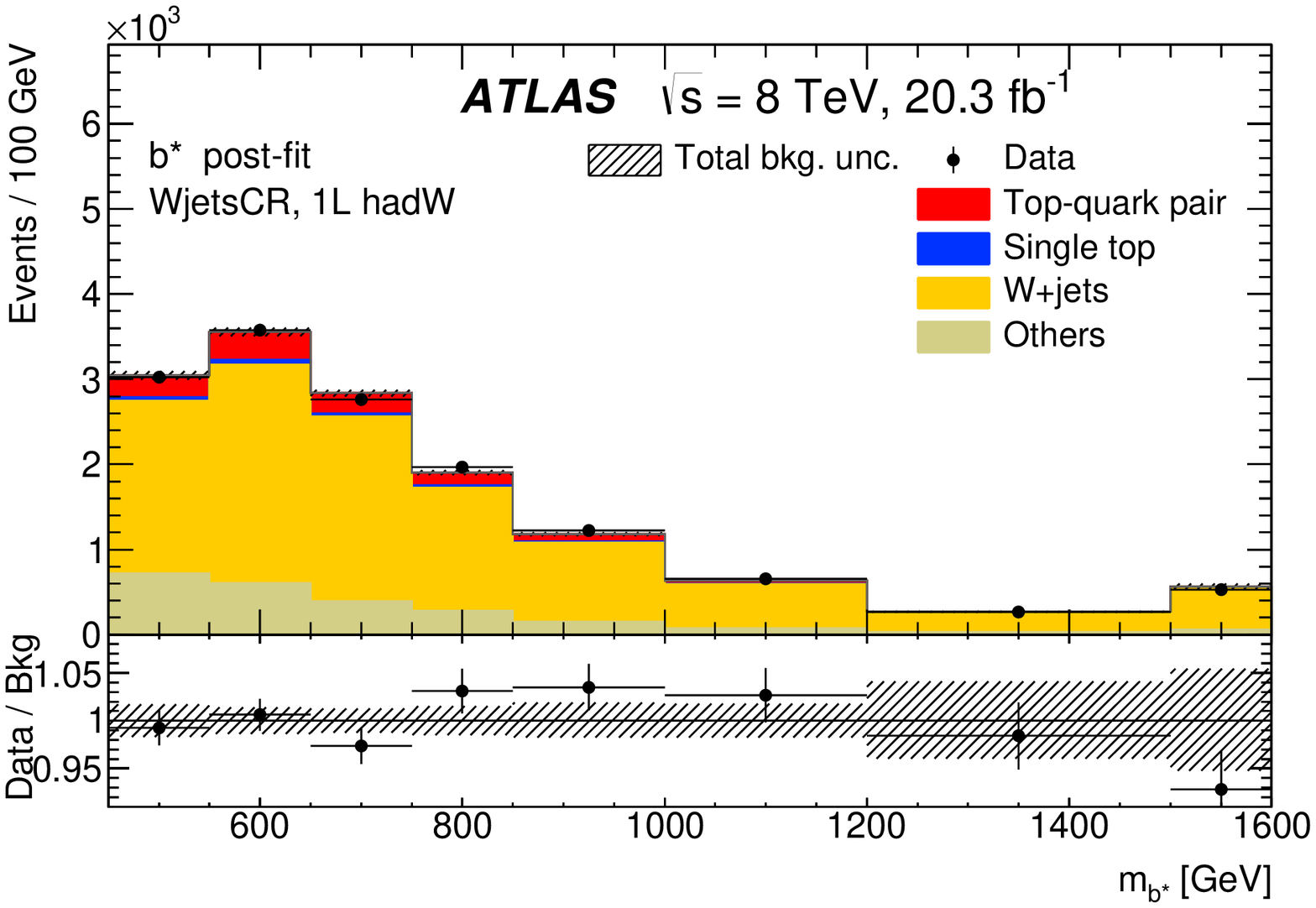} }
\subfigure[]{
\label{F:ttbar3CR_m_tot_only_FWjet_hadTop}
\includegraphics[width=0.48\textwidth]{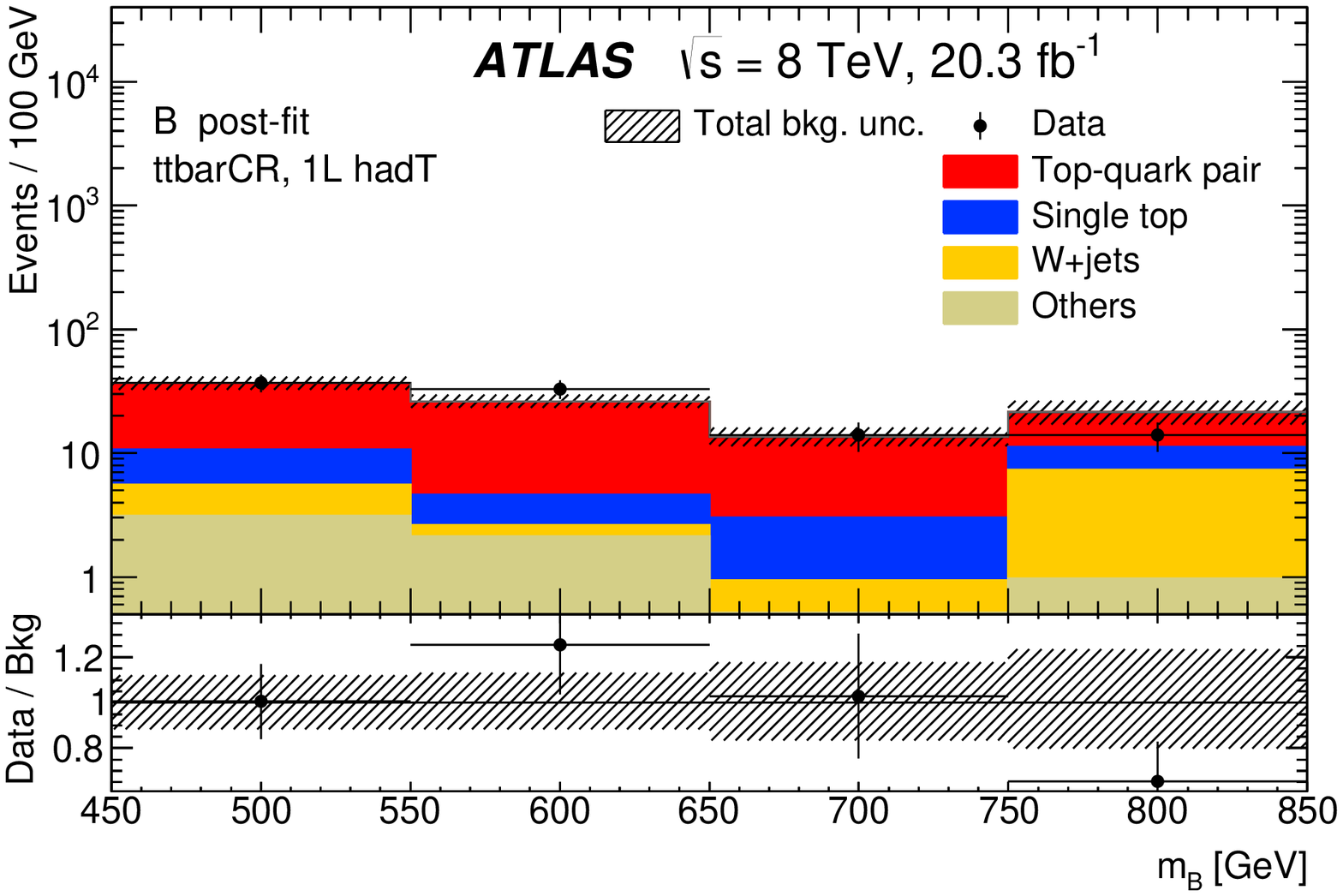} }
\subfigure[]{
\label{F:ttbar3CR_m_tot_only_FWjet_hadW}
\includegraphics[width=0.48\textwidth]{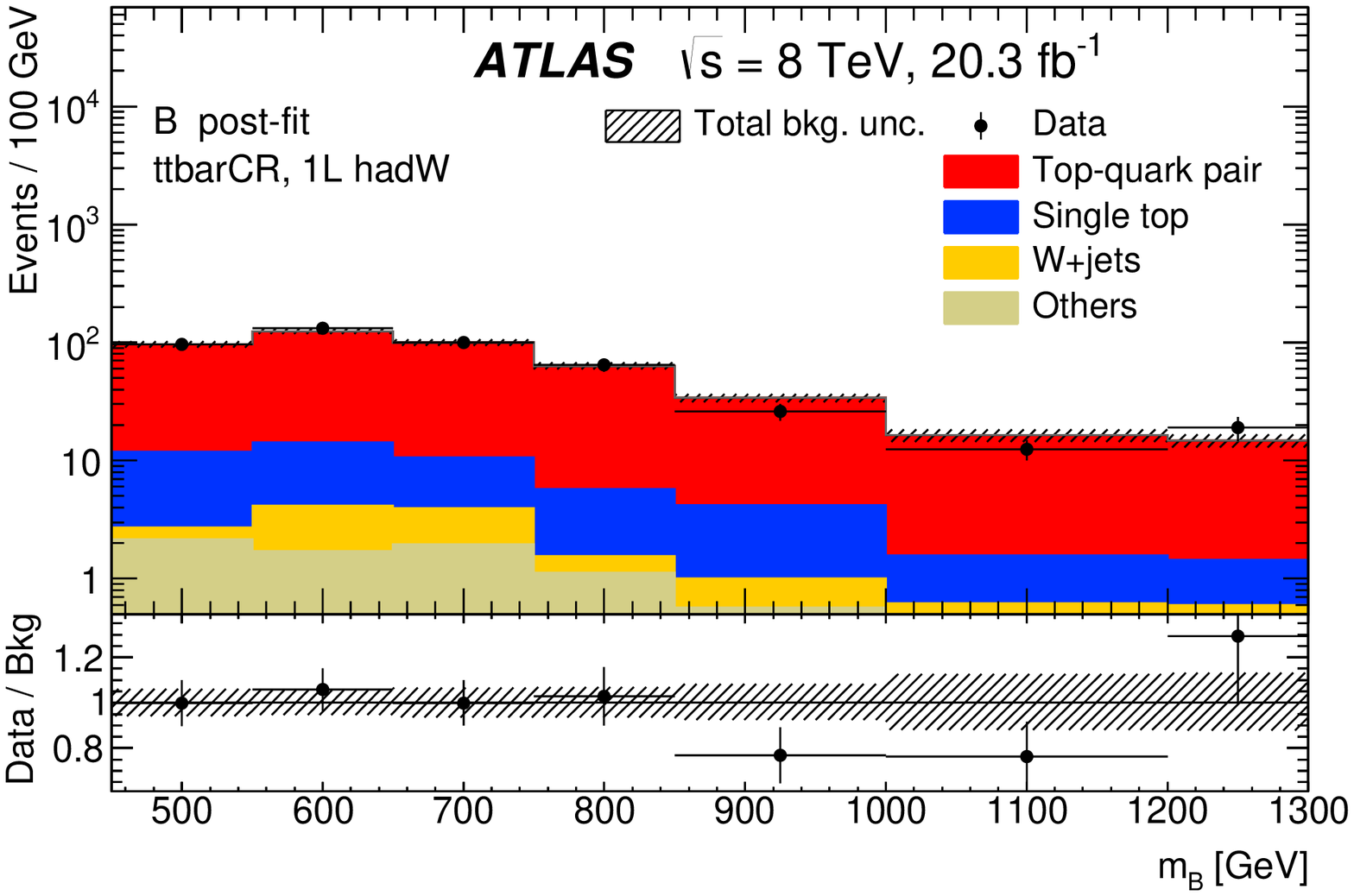} }

\caption{Distributions of the reconstructed mass in the \bstar\ (top row) and \Bprime\ (bottom row) selection of the single-lepton analysis for the
SM background processes and data events in the \subref{F:WjetsCR_m_tot_only_hadTop}, \subref{F:ttbar3CR_m_tot_only_FWjet_hadTop} \textit{1L hadT} and
\subref{F:WjetsCR_m_tot_only_hadW}, \subref{F:ttbar3CR_m_tot_only_FWjet_hadW} \textit{1L hadW} $W$+jets control region (top row) and \ttbar\ control region
(bottom row). 
The background processes are normalised to their post-fit normalisation. The uncertainty band includes statistical and all systematic uncertainties. The first and last bin include the underflow and overflow, respectively. The binning of the figures is the same as used in the binned likelihood fit function. 
\label{fig:lplus_jets_ttbar3CR_WCR_m_tot}}
\end{center}
\end{figure}

Figure~\ref{fig:lplus_jets_ttbar3CR_WCR_m_tot} presents the distributions of the discriminant in the single-lepton channel
(reconstructed mass) together with their statistical and systematic uncertainties in the \bstar\ $W$+jets control regions (top row) and in the \Bprime\ $\ttbar$ control regions (bottom row) for the \textit{1L hadT} category (left side) and the \textit{1L hadW} category (right side).

\begin{table}[htbp]
\centering
\begin{tabular}{l
S[table-format=3.2(2),table-align-uncertainty=true]
S[table-format=4.2(2),table-align-uncertainty=true]
S[table-format=4.3(3),table-align-uncertainty=true]
S[table-format=4.3(3),table-align-uncertainty=true]
}
\toprule
Process & \multicolumn{1}{c}{{\it 1L hadT}} & \multicolumn{1}{c}{{\it 1L hadW}} & \multicolumn{1}{c}{{\it 2L 1jet 1tag}} & \multicolumn{1}{c}{{\it 2L 2jet 1tag}} \\
\midrule
$t\bar{t}$                               &    483     \pm    27     &   746     \pm    34     &   2480    \pm    40     &   1457     \pm    34     \\
Single $t$                               &     94     \pm    13     &   125     \pm    13     &   276     \pm    23     &     81     \pm     9     \\
$W$+jets                                 &    220     \pm    40     &   152     \pm    27     &  \multicolumn{1}{c}{--}  &  \multicolumn{1}{c}{--}  \\
$WW$, $WZ$, $ZZ$, $Z$+jets               &     22     \pm     7     &    30     \pm    11     &   15     \pm     5     &     5     \pm     4     \\
Non-prompt lepton                        &     18     \pm    11     &    28     \pm    14     &    6     \pm     4     &     1.9   \pm     1.5   \\
\hline
Signal $\mBprime=\SI{500}{\GEV}$         &     12.0   \pm     1.9   &    10.4   \pm     1.6   &   13.8   \pm     0.9   &     4.1   \pm     0.4   \\
Signal $\mBprime=\SI{600}{\GEV}$         &     22.4   \pm     2.2   &    13.4   \pm     1.5   &    7.9   \pm     0.4   &     2.34  \pm     0.25  \\
Signal $\mBprime=\SI{700}{\GEV}$         &     16.7   \pm     1.3   &    10.6   \pm     0.8   &    4.06  \pm     0.25  &     1.33  \pm     0.11  \\
Signal $\mBprime=\SI{800}{\GEV}$         &     11.1   \pm     0.7   &     7.3   \pm     0.5   &    2.35  \pm     0.14  &     0.69  \pm     0.06  \\
Signal $\mBprime=\SI{1000}{\GEV}$        &      3.62  \pm     0.29  &     2.78  \pm     0.16  &    0.64  \pm     0.04  &     0.205 \pm     0.019 \\
Signal $\mBprime=\SI{1200}{\GEV}$        &      1.40  \pm     0.09  &      1.09  \pm     0.07 &    0.188 \pm     0.014 &     0.073 \pm     0.007 \\
\midrule
Total bkg.                               &    835     \pm    34     &   1081     \pm    35     &   2780     \pm    40     &   1544     \pm    35     \\
\midrule
Data                                     &    856                   &   1018                   &   2760                   &   1548                   \\
\bottomrule
\end{tabular}

\caption{Event yields in the signal regions for the $\Bprime$ analysis after the fit of the background-only hypothesis (as explained in Section~\ref{sec:res}).
The uncertainties include statistical and systematic uncertainties.
The uncertainties on the background composition are much larger than the uncertainty on the sum of the backgrounds, which is strongly constrained by the data.
The numbers of the $\Bprime$ signal events are evaluated using the respective theory cross-sections and branching ratios for $\lambda=2$.}
\label{TABLE-EVENT-YIELD-Bprime}
\end{table}

\begin{table}[htbp]
\centering
\begin{tabular}{l
S[table-format=4.1(1),table-align-uncertainty=true]
S[table-format=4.1(1),table-align-uncertainty=true]
S[table-format=4.2(2),table-align-uncertainty=true]
S[table-format=4.2(2),table-align-uncertainty=true]
}
\toprule
Process & \multicolumn{1}{c}{{\it 1L hadT}} & \multicolumn{1}{c}{{\it 1L hadW}} & \multicolumn{1}{c}{{\it 2L 1jet 1tag}} & \multicolumn{1}{c}{{\it 2L 2jet 1tag}} \\
\midrule
$t\bar{t}$                               &   1860     \pm    60     &   3390     \pm   120     &   1160     \pm    40     &   1409     \pm    33     \\
Single $t$                               &    368     \pm    33     &    520     \pm    40     &    253     \pm    26     &     97     \pm    10     \\
$W$+jets                                 &   1360     \pm   140     &   1290     \pm   120     &  \multicolumn{1}{c}{--}  &  \multicolumn{1}{c}{--}  \\
$WW$, $WZ$, $ZZ$, $Z$+jets               &    230     \pm   100     &    180     \pm    50     &     13     \pm     4     &      3.0   \pm     1.6   \\
Non-prompt lepton                        &     60     \pm    40     &    100     \pm    40     &      6     \pm     4     &      2.0   \pm     1.4   \\
\midrule
Signal $\mbstar=\SI{600}{\GEV}$          &   5800     \pm   270     &   4890     \pm   230     &    800     \pm    50     &    493     \pm    27     \\
Signal $\mbstar=\SI{800}{\GEV}$          &   2240     \pm    90     &   1640     \pm    70     &    215     \pm    14     &    142     \pm     7     \\
Signal $\mbstar=\SI{1000}{\GEV}$         &    661     \pm    30     &    591     \pm    26     &     55.0   \pm     3.4   &     44.1   \pm     2.5   \\
Signal $\mbstar=\SI{1200}{\GEV}$         &    209     \pm    10     &    196     \pm    10     &     16.2   \pm     1.2   &     14.2   \pm     1.0   \\
Signal $\mbstar=\SI{1400}{\GEV}$         &     68.5   \pm     3.2   &     63.4   \pm     3.1   &      4.8   \pm     0.4   &      4.20  \pm     0.34  \\
Signal $\mbstar=\SI{1600}{\GEV}$         &     23.8   \pm     1.2   &     21.2   \pm     1.0   &      1.41  \pm     0.15  &      1.35  \pm     0.14  \\
Signal $\mbstar=\SI{1800}{\GEV}$         &      8.7   \pm     0.5   &      7.3   \pm     0.4   &      0.49  \pm     0.08  &      0.48  \pm     0.08  \\
\midrule
Total bkg.                               &   3880     \pm    70     &   5480     \pm    70     &   1436     \pm    33     &   1511     \pm    34     \\
\midrule
Data                                     &   3933                   &   5380                   &   1448                   &   1502                   \\
\bottomrule
\end{tabular}

\caption{Event yields in the signal regions for the $\bstar$ analysis after the fit of the background-only hypothesis (as explained in Section~\ref{sec:res}).
The uncertainties include statistical and systematic uncertainties. 
The uncertainties on the background composition are much larger than the uncertainty on the sum of the backgrounds, which is strongly constrained by the data.
The numbers of the $\bstar$ signal events are evaluated using the respective theory cross-sections and branching ratios.
The couplings are assumed to be $\fL=\fR=1$.}
\label{TABLE-EVENT-YIELD-bstar}
\end{table}

\subsection{Dilepton event selection and background estimation}
\label{sec:seldl}
Events in the dilepton final state are required to have exactly one electron and one muon with opposite charge as well as one or two central small-$R$ jets ($|\eta|<2.5$), exactly one of which is required to be $b$-tagged.
Since same-flavour $Z\to \ell \ell$ ($\ell=e,\mu$) events are already suppressed by the opposite-flavour lepton requirement, no $\met$ requirement is applied. 
For the $\bstar$ signal, it was found that the angular distance between the jet and the lepton is smaller than that in background process events, so an additional requirement is made on the smallest value of angle $\phi$ between the leading small-$R$ jet ($j_0$) and the leptons, $\Delta \phi_\mathrm{min}(\ell,j_0)<0.9$. 

For the $\Bprime$ signal, events are required to have exactly one central $b$-tagged jet from the top-quark decay and one small-$R$ jet with $1.5 <|\eta| < 4.5$ from the light quark. If this additional jet is in the range of $|\eta|<2.5$, then it is required not to be $b$-tagged. 
Events are also allowed to contain up to one additional untagged jet with $|\eta|$<1.5.

Events with one central jet and two central jets have different background compositions and are considered as separate signal regions {\it 2L 1jet 1tag} and {\it 2L 2jet 1tag}, respectively. 
A $\ttbar$ control region {\it 2L 2jet 2tag} is defined by requiring exactly two central $b$-tagged small-$R$ jets.
The contamination by $\Bprime$ and $\bstar$ signals with $m=\SI{1000}{\GEV}$ is at most $0.3\%$.

The main background processes with a dilepton signature arise from top-quark pair, single top-quark and diboson production, with smaller contributions from non-prompt lepton and  $Z$+jets events. 
The $Z$+jets background, which arises from $Z$-boson decays to tau leptons that subsequently decay into an electron and a muon, is taken from simulation samples. 
The non-prompt lepton background process arises from events where one or two jets are misidentified as leptons.
This background includes mainly multijet processes as well as processes with a single-lepton final state ($\ttbar$ single-lepton channel, single top-quark other than $\Wt$-channel, $W+$jets).
It is estimated with the data-driven matrix method~\cite{ATLAS-CONF-2014-058} as described in Section~\ref{sec:sellj}.
The event yields in each of the signal regions after fitting the background-only hypothesis to data are shown in Table~\ref{TABLE-EVENT-YIELD-Bprime} and~\ref{TABLE-EVENT-YIELD-bstar}. 
For each process the post-fit normalisation is compatible with the pre-fit normalisation.

\begin{figure}
\centering
\subfigure[]{
 \label{subfig:signalallcutbprimeLH1CR:leppt:m} 
 \includegraphics[width=0.48\textwidth]{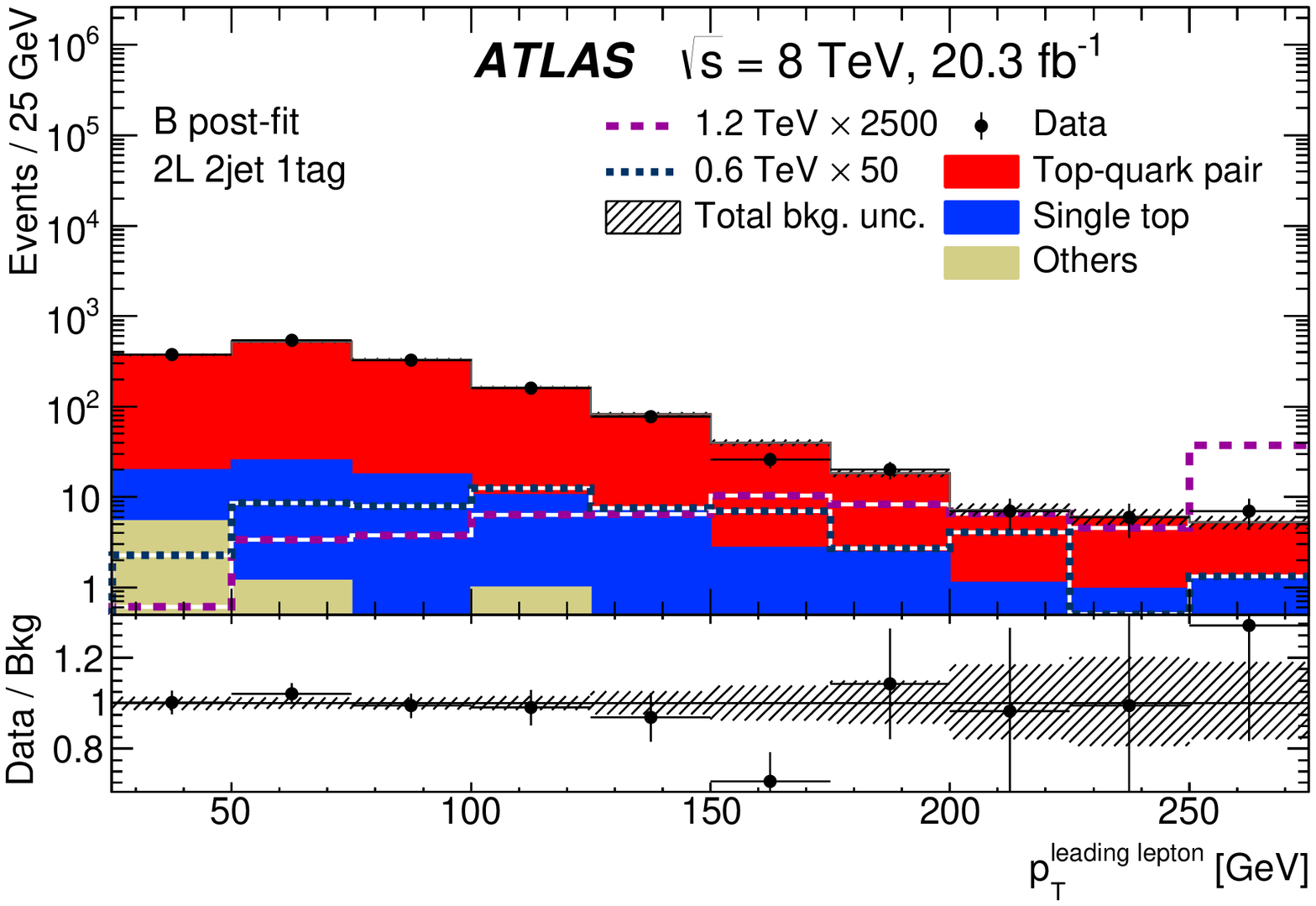}
 }
\subfigure[]{
 \label{subfig:signalallcutbstarLH1:leppt:m} 
 \includegraphics[width=0.48\textwidth]{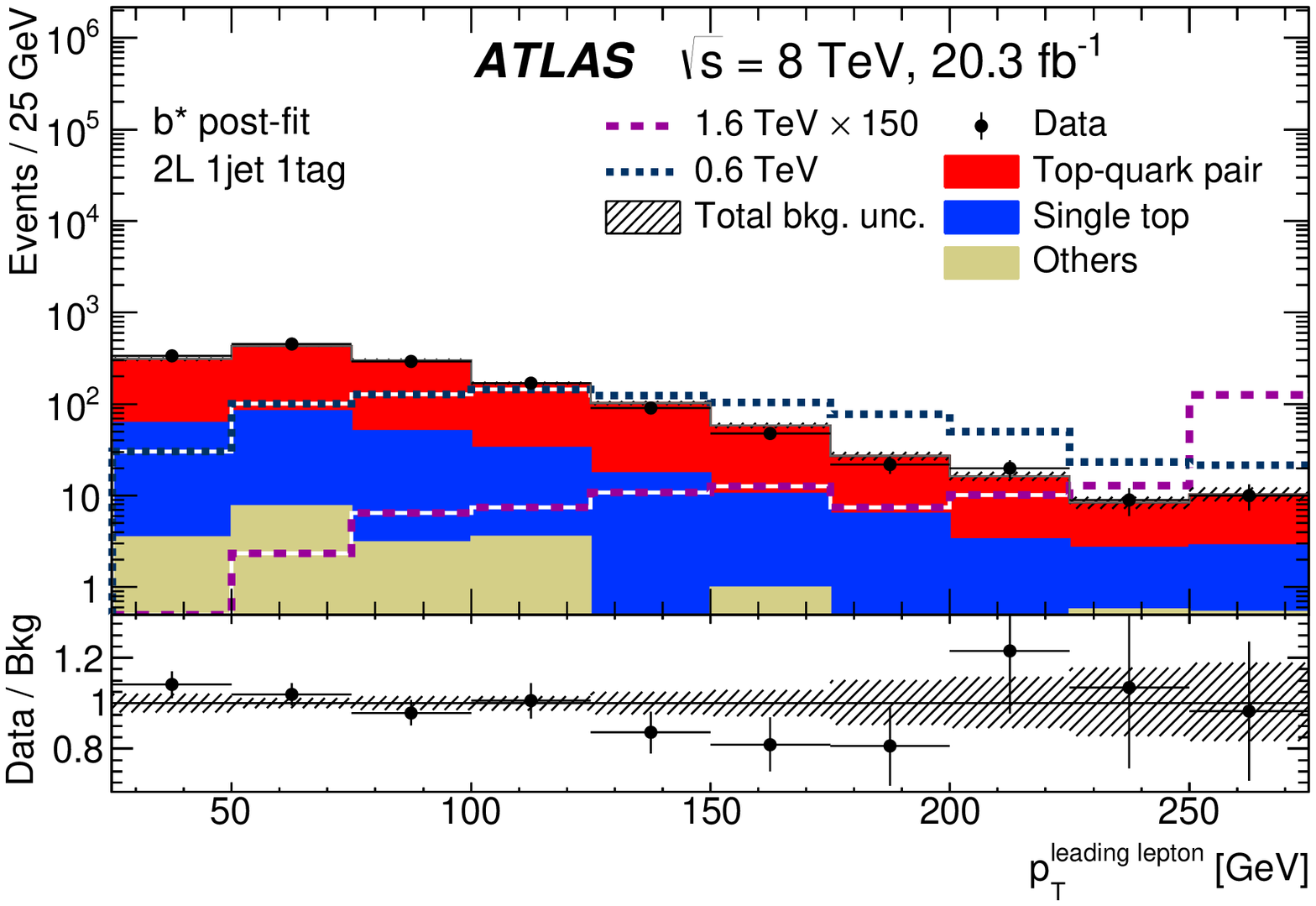}
 }
\subfigure[]{
 \label{subfig:signalallcutbprimeLH1:met:m} 
 \includegraphics[width=0.48\textwidth]{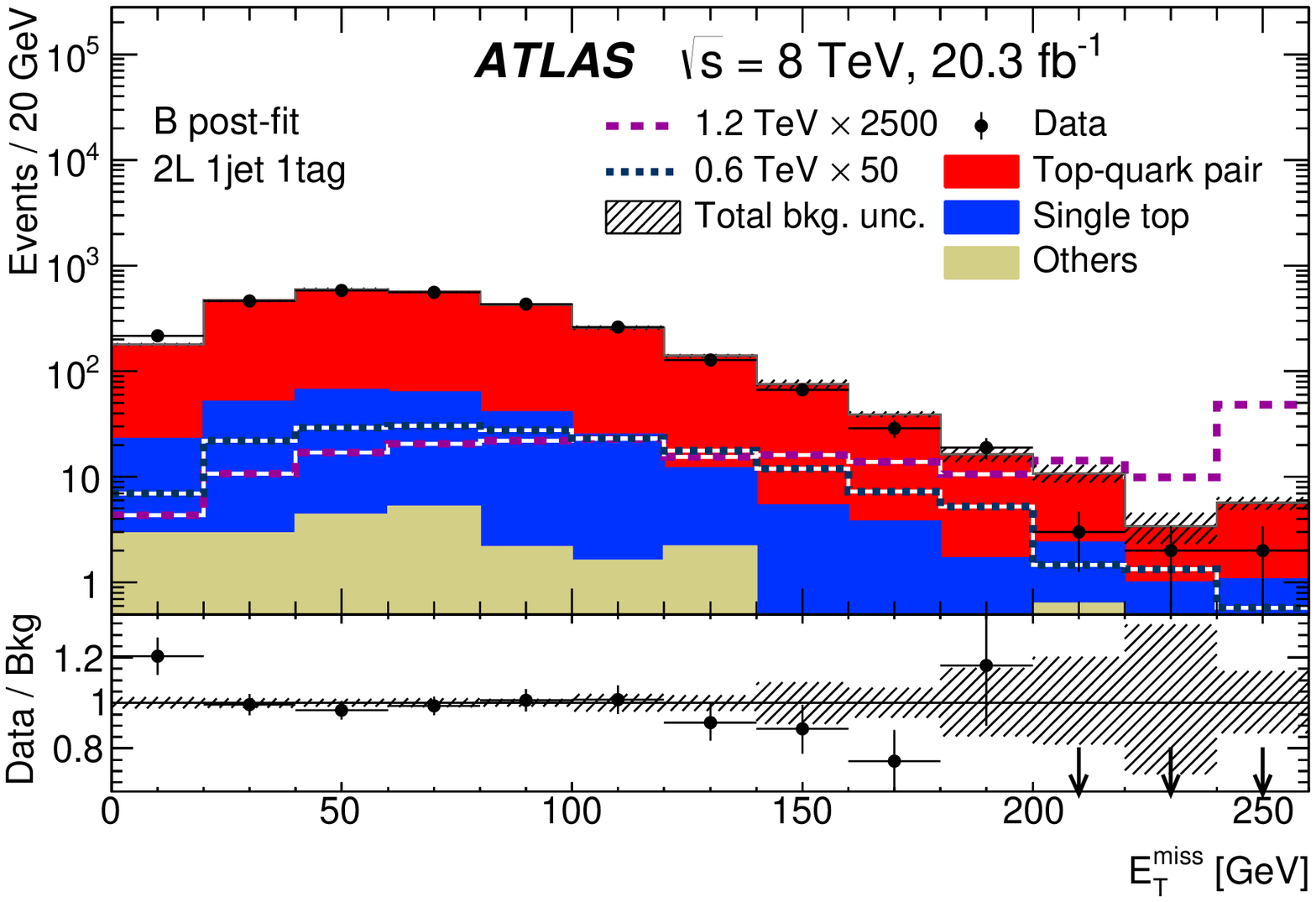}
 }
\subfigure[]{
 \label{subfig:signalallcutbstarLH2:met:m} 
 \includegraphics[width=0.48\textwidth]{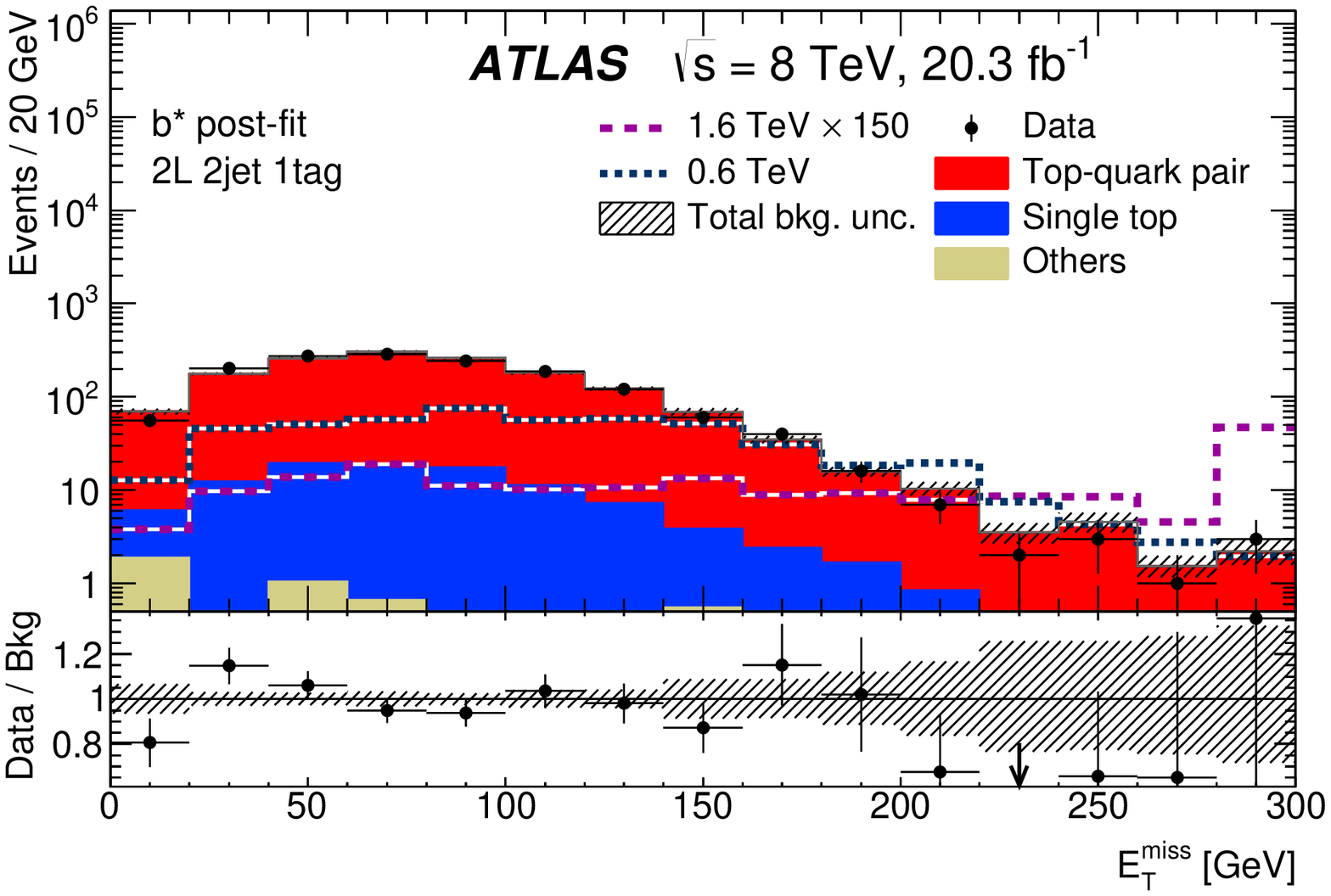}
 }
\caption{Distributions of kinematic variables in different signal regions of the dilepton selection for SM background processes and data, overlaid with two corresponding signal distributions:
\subref{subfig:signalallcutbprimeLH1CR:leppt:m} leading lepton $\pt$ in the {\it 2L 2jet 1tag} region after the $\Bprime$ selection and 
\subref{subfig:signalallcutbstarLH1:leppt:m} in the {\it 2L 1jet 1tag} region after the $\bstar$ selection, 
\subref{subfig:signalallcutbprimeLH1:met:m} $\met$ in the  {\it 2L 1jet 1tag} region after the $\Bprime$ selection and 
\subref{subfig:signalallcutbstarLH2:met:m} in the {\it 2L 2jet 1tag} region after the $\bstar$ selection.
The signal distributions are scaled to the theory cross-sections multiplied by the indicated factor for better visibility and the background processes are normalised to their post-fit normalisation.
The uncertainty band includes statistical and all systematic uncertainties. 
The last bin includes the overflow.
\label{fig:dilepkinstarCRmain}}
\end{figure}

Selected kinematic distributions, such as the $\pt$ of the leading lepton and the $\met$, for the dilepton events in the signal regions {\it 2L 1jet 1tag} and {\it 2L 2jet 1tag} are shown in Figure~\ref{fig:dilepkinstarCRmain}.
The background processes are normalised to post-fit event yields and their sum models the data
well. The transverse mass ($m_\mathrm{T}$) of the system formed by the leptons, the leading jet and the $\met$ is used as the final discriminating distribution. 

\begin{figure}
\centering
\subfigure[]{
 \label{subfig:mtBprime1j} 
 \includegraphics[width=0.48\textwidth]{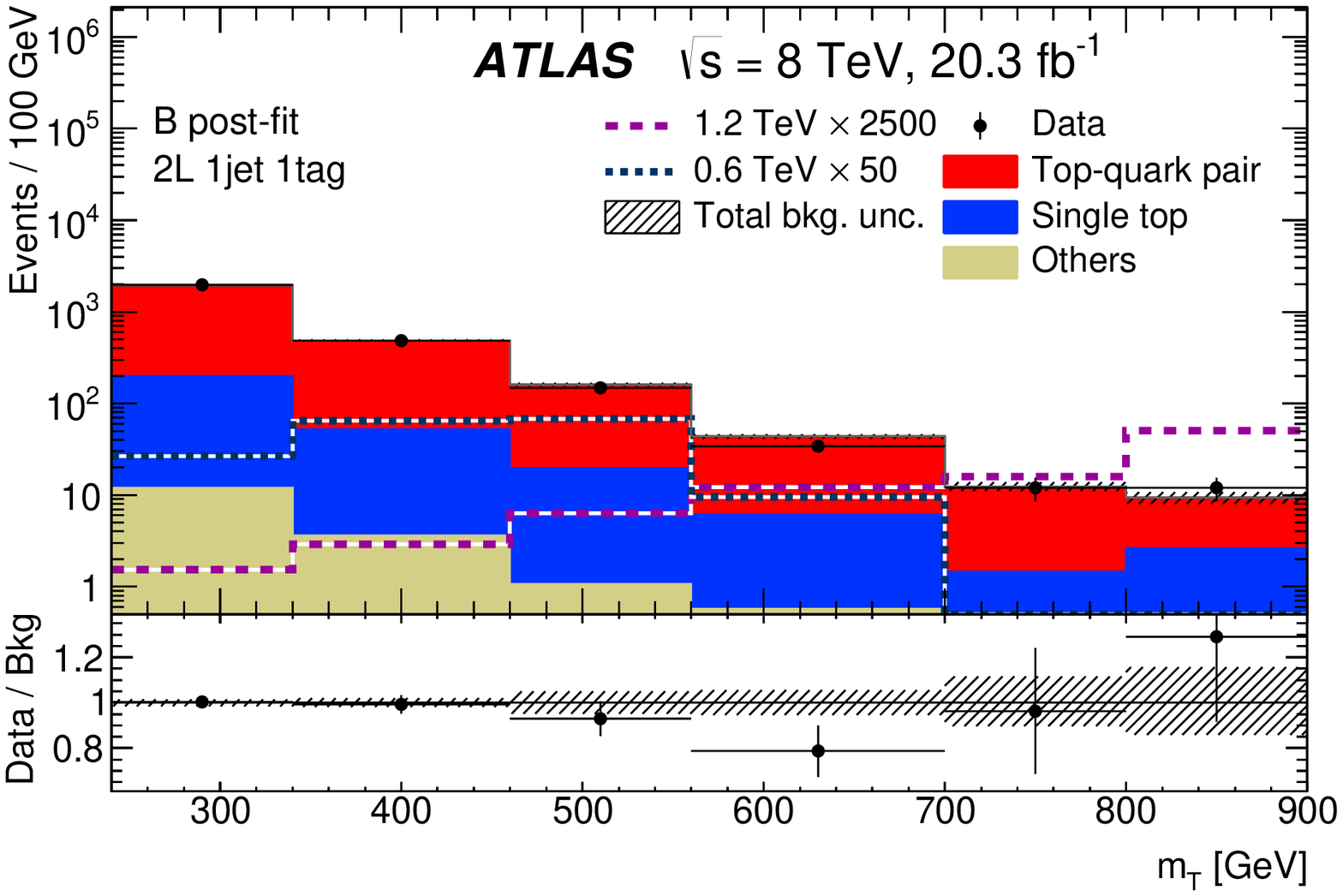}
 }
\subfigure[]{
 \label{subfig:mtBprime2j} 
 \includegraphics[width=0.48\textwidth]{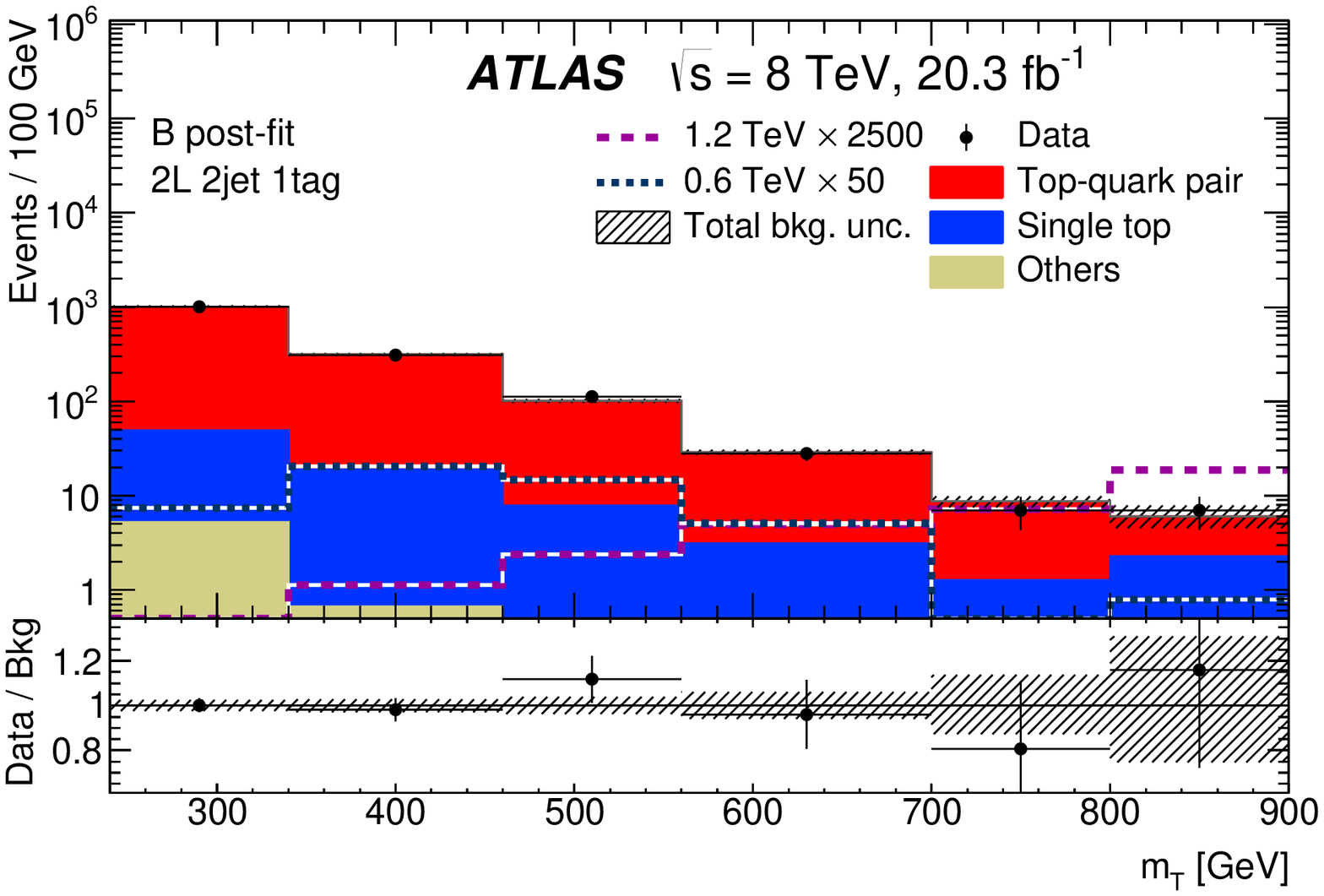}
 }
\subfigure[]{
 \label{subfig:mtBprime2j2t} 
 \includegraphics[width=0.48\textwidth]{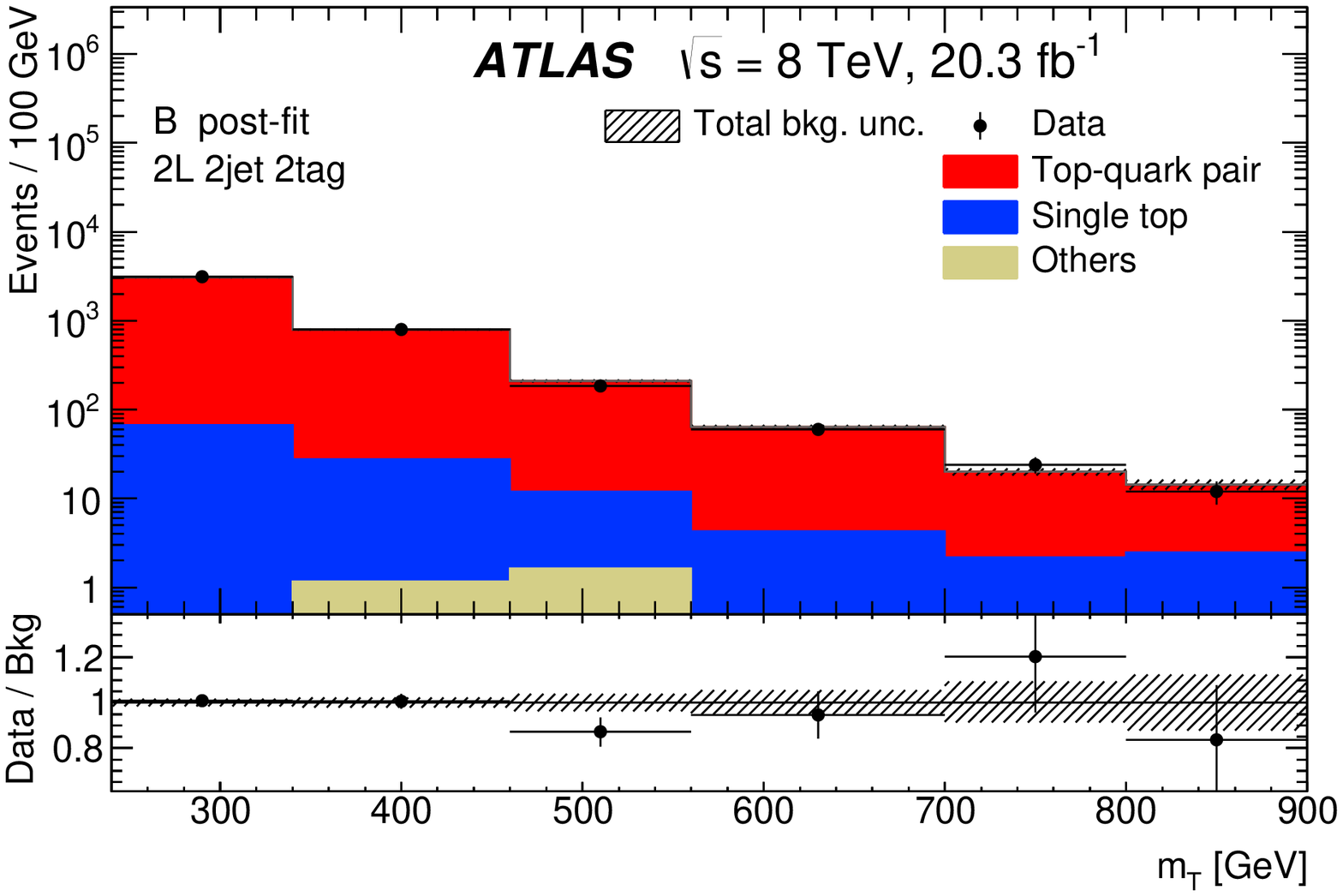}
 }
\caption{Distributions of the discriminant variable, transverse mass, $m_{\rm T}$, in the dilepton channel for data and SM background processes in the \subref{subfig:mtBprime1j} {\it 2L 1jet 1tag} signal region, \subref{subfig:mtBprime2j} {\it 2L 2jet 1tag} signal region and \subref{subfig:mtBprime2j2t} {\it 2L 2jet 2tag} control region using the $\Bprime$ selection.
The two $\Bprime$ signal distributions that are included for the signal region distributions are scaled to the theory cross-sections multiplied by the indicated factor for better visibility. 
The background processes are normalised to their post-fit normalisation.
The uncertainty band includes statistical and all systematic uncertainties. 
The first and the last bin contain the underflow and overflow, respectively.
The binning of the figures is the same as used in the binned likelihood fit function. 
\label{fig:dilepBprimemt} }
\end{figure}

\begin{figure}
\centering
\subfigure[]{
 \label{subfig:mtbstar1j} 
 \includegraphics[width=0.48\textwidth]{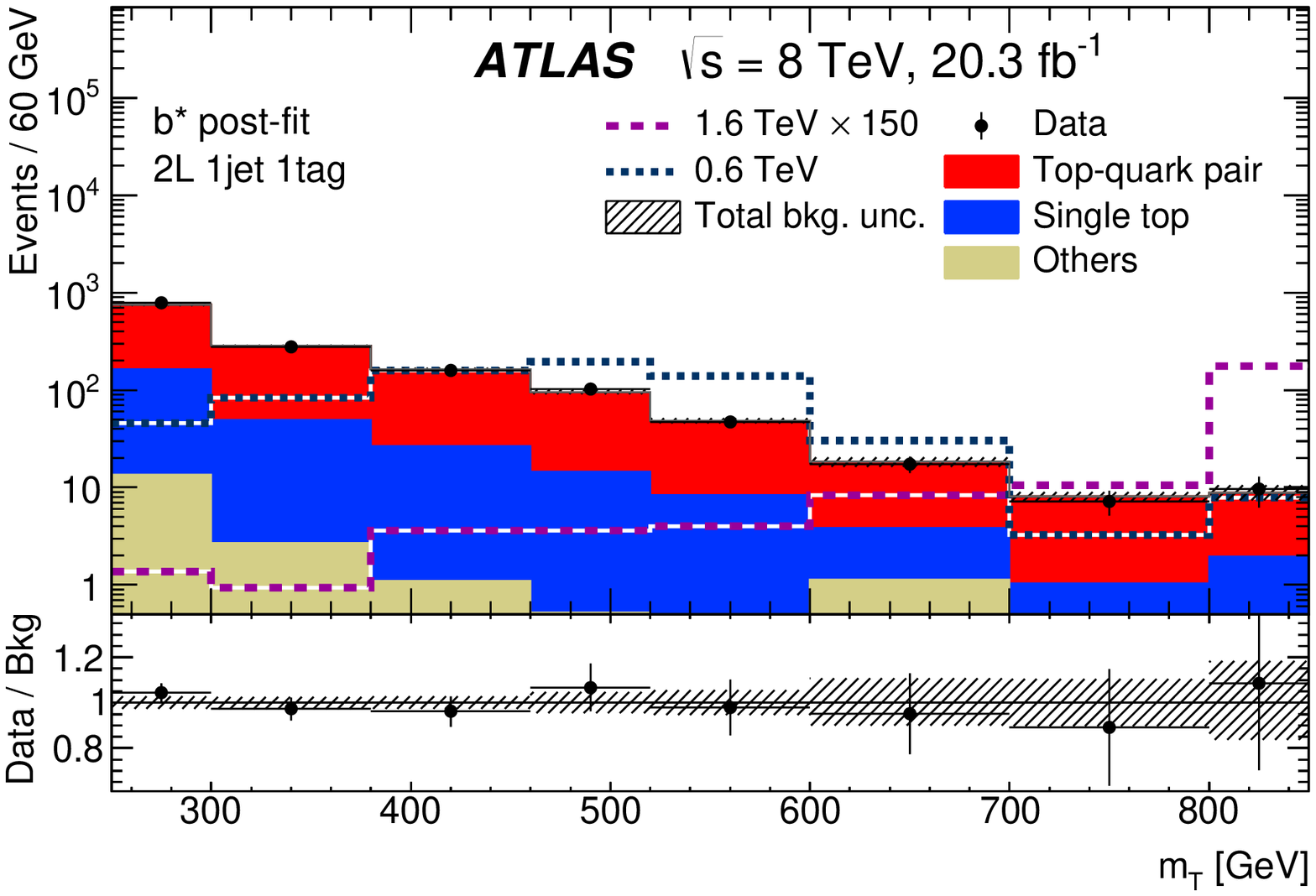}
 }
\subfigure[]{
 \label{subfig:mtbstar2j} 
 \includegraphics[width=0.48\textwidth]{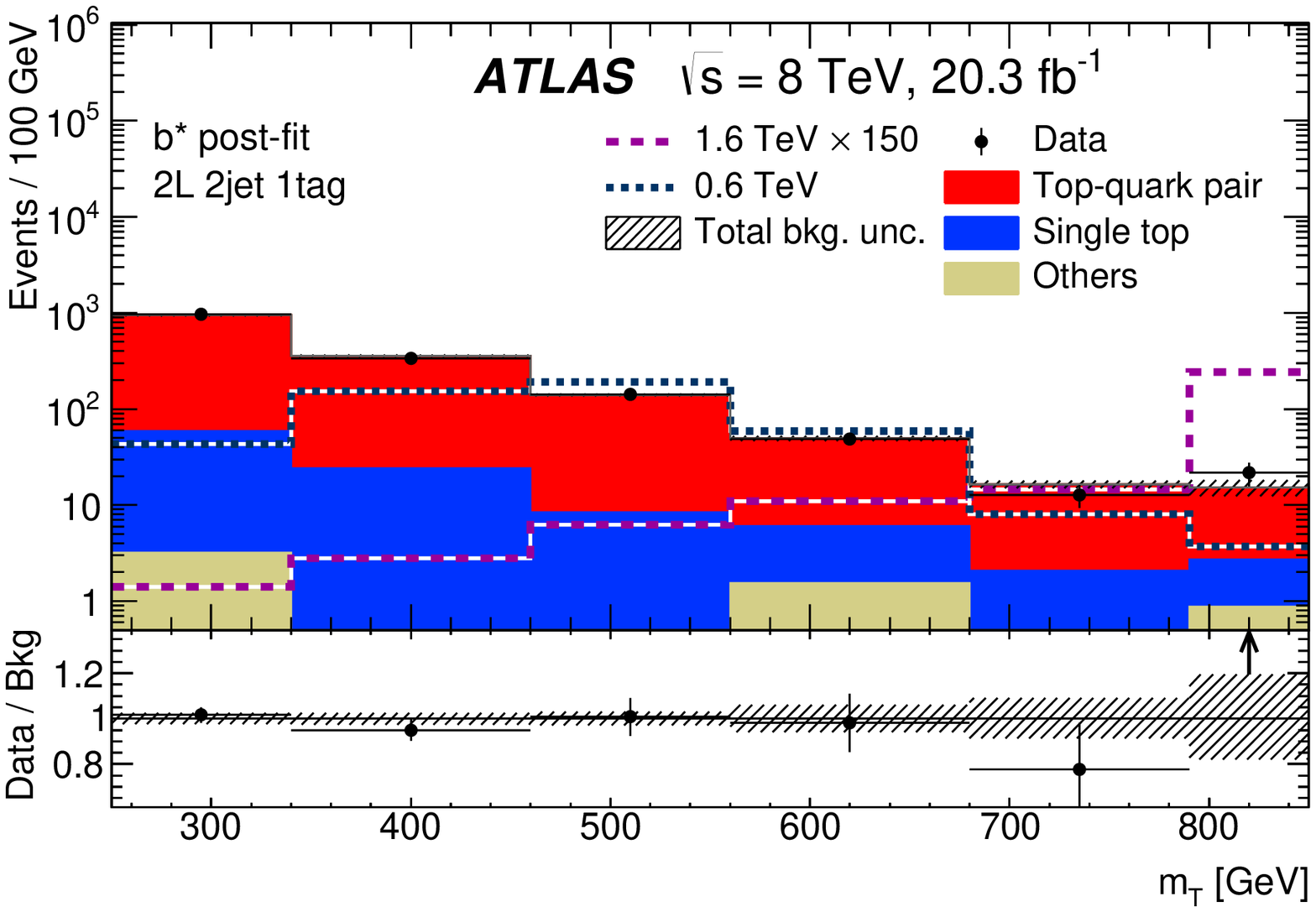}
 }
\subfigure[]{
 \label{subfig:mtbstar2j2t} 
 \includegraphics[width=0.48\textwidth]{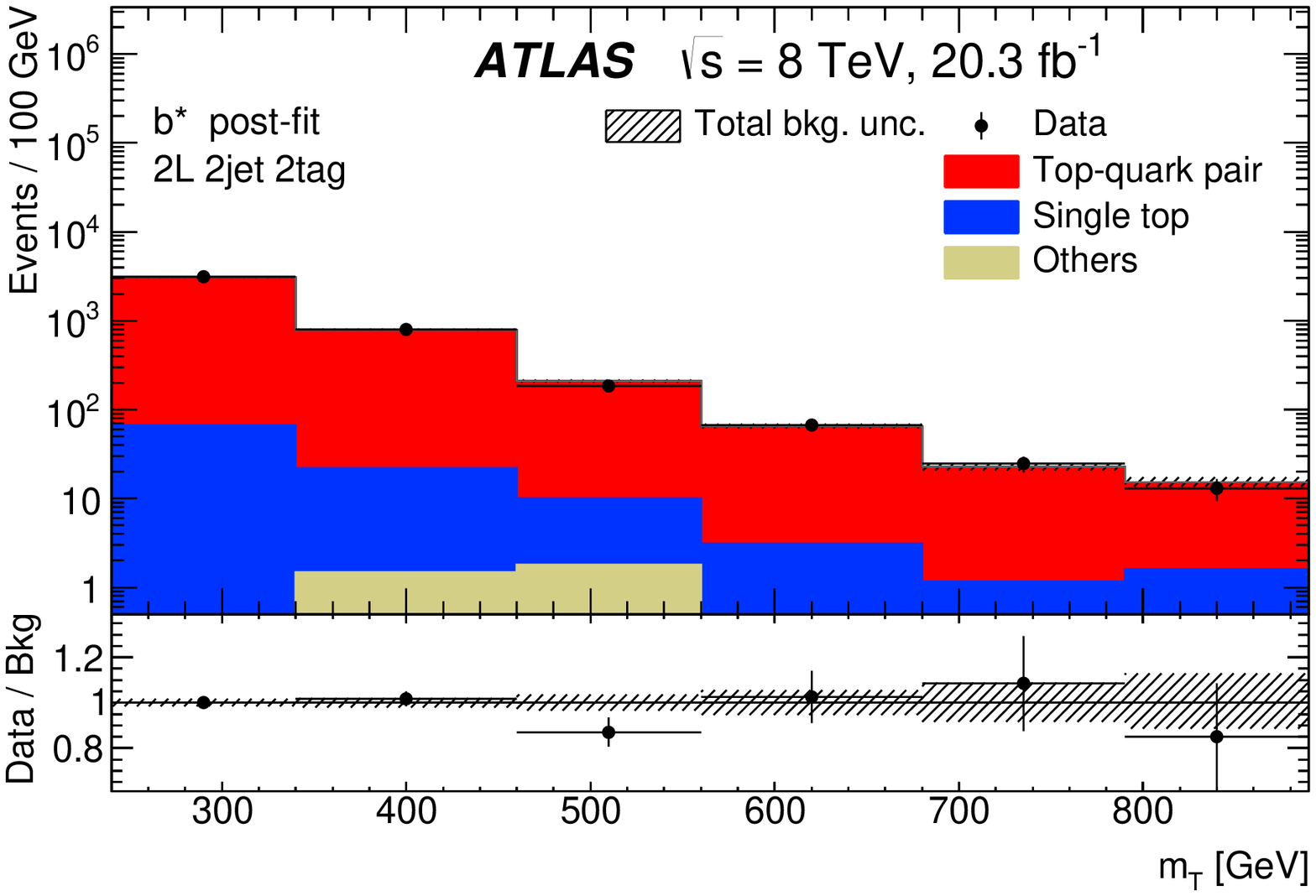}
 }
\caption{Distributions of the discriminant variable, transverse mass, $m_{\rm T}$, in the dilepton channel for data and SM background processes in the \subref{subfig:mtbstar1j} {\it 2L 1jet 1tag} signal region, \subref{subfig:mtbstar2j} {\it 2L 2jet 1tag} signal region and \subref{subfig:mtbstar2j2t} {\it 2L 2jet 2tag} control region using the $\bstar$ selection.
For the signal region distributions, the two included $\bstar$ signal distributions are scaled to the theory cross-sections multiplied by the indicated factor for better visibility. 
The background processes are normalised to their post-fit normalisation.
The uncertainty band includes statistical and all systematic uncertainties. 
The first and the last bin contain the underflow and overflow, respectively.
The binning of the figures is the same as used in the binned likelihood fit function. 
\label{fig:dilepbstarmt} }
\end{figure}

The distributions of $m_\mathrm{T}$ for the different $\Bprime$ and $\bstar$ signal and control regions are shown in Figure~\ref{fig:dilepBprimemt} and Figure~\ref{fig:dilepbstarmt}, respectively.
The distributions show the bin widths used in the statistical analysis.
The control region distributions for various kinematic variables as well as the event yields show good agreement between data and MC simulation.

\section{Systematic uncertainties}
\label{sec:sys}

Systematic uncertainties affect the yield and acceptance estimates as well as the shape of the discriminant distributions of the signal and background processes. They are included as nuisance parameters in the statistical analysis and both normalisation and shape variations are considered (except for the theoretical cross section uncertainty and the PDF uncertainties).
Sources of uncertainty include the modelling of the detector, properties of reconstructed objects, theoretical modelling of the signals
and backgrounds, as well as the uncertainty of the prediction arising from the limited size of the simulated event samples.

\subsection{Experimental uncertainties}
Experimental sources of systematic uncertainty arise from the reconstruction and measurement of jets~\cite{Aad:2014bia,ATLAS-CONF-2015-017,ATLAS-CONF-2012-065,Aad:2012ag,ATLAS-CONF-2013-083}, leptons~\cite{Aad:2014fxa,Aad:2014zya} and $\met$~\cite{Aad:2012re}. 
The impact of the uncertainty in the jet-energy scale (JES) is evaluated for both the small-$R$ and large-$R$ jets. 
For large-$R$ jets, additional uncertainties associated with the scale and resolution of the jet mass are included.
The $\met$ has additional uncertainty contributions due to the modelling of energy deposits not associated with any reconstructed object.
Further uncertainties arise from  the lepton trigger, reconstruction, energy scale and resolution as well as from the lepton
and jet identification efficiency modelling. Uncertainties on the $b$-tagging efficiency and mistag rates are estimated from data~\cite{ATLAS-CONF-2014-004}.

The largest detector-specific uncertainties for the single-lepton channel arise from large-$R$ jet-energy scale uncertainties (10--18\% effect on the predicted background yield), large $R$-jet energy and mass resolution uncertainties (2--5\%), small-$R$ jet-energy scale uncertainties (3--5\%) and the $b$-tagging uncertainties (1--3\%), depending on the signal region and analysis selection.
The expected yield of events with non-prompt leptons is subject to uncertainties in the real and non-prompt lepton efficiencies and they sum to a 2--18\% uncertainty on the total SM background in the defined signal regions.

For the dilepton channel, the relative impact of a particular systematic uncertainty is similar for the signal and background and for the different signal regions.
Significant variations on the acceptance come from the $b$-tagging uncertainty (2--5\%), electron identification uncertainty (2--3\%) and from JES uncertainties related to pile-up (1--2\%).

\subsection{Modelling of theoretical uncertainties}

Theoretical uncertainties are evaluated for the signal as well as for the background processes. The evaluation of the PDF uncertainty follows the PDF4LHC prescription~\cite{Botje:2011sn} using three different PDF sets ({\scshape CT10}, {\scshape MSTW2008nlo68cl} and {\scshape NNPDF20}~\cite{Ball:2010de}).
The uncertainty on top-quark pair production due to initial- and final-state radiation is evaluated using the {\scshape Alpgen} LO generator, with the CTEQ6L1 PDF set and interfaced with {\scshape Pythia} v6.426.
The renormalisation scale associated with the strong coupling $\alpha_\mathrm{S}$ in the matrix-element calculation is varied by factors of $2$ and $0.5$, and at the same time the amount of radiation is decreased or increased in a range compatible with data by choosing the appropriate variation of the {\scshape Pythia} Perugia 2012 tune~\cite{Skands:2010ak}.

The uncertainty on the top-quark pair and single top-quark $\Wt$ processes due to the choice of generator and parton shower is evaluated by comparing the nominal simulation samples to {\scshape Powheg-Box} and {\scshape MC@NLO} v4.06~\cite{Frixione:2002ik,Frixione:2003ei} samples interfaced with {\scshape Herwig} v6.52.

For background estimates based on simulations, the largest sources of uncertainty in the single-lepton analysis are due to varying the parameters controlling the initial-state radiation model (2.5--5\%), the PDF sets (1--3\%) and renormalisation and factorisation scales (1--7\%).
Differences for heavy-flavour and light-flavour components of the $W$+jets background processes in the signal and control regions
lead to a systematic uncertainty of 4--5\%.

For the dilepton analysis, the initial- and final-state radiation uncertainty has an impact on the selection efficiency of 5--17\%.
The impact of the other theoretical uncertainties on the total background rate is 1--8\%, depending on the signal region and signal or background sample.

The normalisation of the $\ttbar$, single-top $t$-channel and $\Wt$ backgrounds has an uncertainty of ${}^{+5.1}_{-5.9}\%$~\cite{Czakon:2013goa,Czakon:2011xx}, ${}^{+3.9}_{-2.2}\%$ and 6.8\%~\cite{Kidonakis:2010ux}, respectively.
The diboson and $W/Z$+jets background processes have an uncertainty of 5\% and 4\%, respectively~\cite{Anastasiou:2003ds}, with an additional 24\% for each additional jet~\cite{Berger:2010zx,Berends:1990ax}. 

The uncertainty on the luminosity of $\pm 2.8\%$~\cite{Aad:2013ucp} affects the normalisation of the signal and the background processes estimated from theory predictions.

\section{Statistical analysis and results}
\label{sec:res}

{%

\def\fLR{f_\mathrm{L,R}}

\newcommand\XXX{\text{\bf?}}

The distributions of $\mBprime$ and $\mbstar$ in the single-lepton channel and $m_\mathrm{T}$ in the dilepton channel are sensitive to resonant production of $\Bprime$ and $\bstar$, respectively.
A binned likelihood fit is performed in order to test for the presence of a signal.
A separate fit is performed for each signal hypothesis, i.e. $\Bprime$/$\bstar$ with fixed mass and couplings.
Let $N_i$ be the observed number of events in bin $i$, $B_i$ the predicted number of background events, $S_i$ the
predicted number of signal events assuming a cross-section of $\SI{1}{\pico\barn}$, and $\mu$ the signal-strength
parameter, which measures the signal cross-section in units of \si{\pico\barn}.
The likelihood function is constructed as the product of Poisson likelihood terms, each of the form $\mathrm{Pois}(N_i;\mu S_i + B_i)$, over all bins of the $\mBprime/\mbstar/\m_\mathrm{T}$ distributions in the signal and control regions.
The one-sided $95\%$ CL upper limit on $\mu$, and therefore the signal cross-section, is constructed in a Bayesian approach\footnote{%
  The posterior density function is sampled using the Bayesian Analysis Toolkit~\cite{caldwell2009bat}, interfaced to RooStats~\cite{cranmer2012histfactory}.
} with a positive semidefinite flat prior for $\mu$.
The expected limits are derived by fitting to the nominal background estimate~\cite{cowan2011asymptotic}.

The likelihood function also depends on a set of nuisance parameters, which encode the adjustments of the expectations due to the systematic uncertainties discussed
in Section~\ref{sec:sys}. The prior for each of the nuisance parameters is taken to be the normal distribution, where a value of $0$ corresponds to the no\-minal prediction, and $\pm1$ to the symmetric variation by one standard deviation.
The uncertainties on the expected numbers of events in each bin ($S_i$, $B_i$) are propagated to $\mu$ through these parameters.
They also allow the control regions to improve the description of the data, and to reduce the impact of systematic uncertainties in the signal region.
The strongest reduction is achieved for the normalisation uncertainty on the $W$+jets background, which
becomes 10\% after the fit, compared to the about 40\% before the fit.
The largest uncertainties are due to the JES, the $b$-tagging, and the background normalisation, where the latter is dominant for small masses.
Table~\ref{tab:syst} shows the normalisation variation from the largest systematic uncertainties before and after the fit.
The posterior distribution of the nuisance parameters after the fit of the background-only ($\mu=0$) hypothesis to the data is used to compute post-fit event yields, to normalise the distributions in the control plots,
and to check for agreement between data and the background prediction in many variables.
Given that the best-fit values of the nuisance parameters were found to be compatible between the individual regions, the nuisance parameters are treated as fully correlated between the regions.

\begin{table}[htbp]
\centering

\begin{tabular}{l
S[table-format=2.1]
S[table-format=2.1]
S[table-format=2.1]
S[table-format=2.1]
}
\toprule
\multirow{2}{*}{} & \multicolumn{2}{c}{$\bstar$} & \multicolumn{2}{c}{$\Bprime$} \\
                              & \multicolumn{1}{c}{Pre-fit [\%]} & \multicolumn{1}{c}{Post-fit [\%]} & \multicolumn{1}{c}{Pre-fit [\%]} & \multicolumn{1}{c}{Post-fit [\%]} \\
\midrule
    Jet uncertainties         & 14.          & 6.5           & 12.          & 6.2  \\
    $b$-tagging uncertainties & 3.3          & 3.0           & 2.8          & 2.5  \\
    Lepton uncertainties      & 1.6          & 1.5           & 1.6          & 1.6  \\
    Fake-lepton uncertainties & 2.6          & 2.4           & 2.9          & 2.6  \\
    Theory uncertainties      &              &               &              &      \\
    \tabitem Top-quark pair   & 3.2          & 1.8           & 9.4          & 3.4  \\
    \tabitem $W$+jets         & 9.1          & 3.6           & 9.6          & 4.9  \\
    \tabitem Single top       & 0.0          & 0.0           & 0.1          & 0.1  \\
    \tabitem Diboson          & 0.5          & 0.5           & 0.2          & 0.2  \\
    \tabitem $Z$+jets         & 0.5          & 0.5           & 0.7          & 0.7  \\
\bottomrule
\end{tabular}
\caption{Relative effects of the systematic uncertainties on the total background estimate in the signal regions, before and after fitting.
The jet uncertainties include large-$R$ jet uncertainties and other jet uncertainties such as jet energy scale and resolution. 
The lower part of the table shows how the theoretical uncertainties on individual backgrounds affect the sum of all backgrounds.
} 
\label{tab:syst}
\end{table}

The post-fit event yields for the $\Bprime$ and $\bstar$ signal regions together with their systematic uncertainties are listed in the Tables~\ref{TABLE-EVENT-YIELD-Bprime} and~\ref{TABLE-EVENT-YIELD-bstar}.
For illustration, the number of events for some signal hypotheses are also presented, each normalised to the predicted cross-section times branching ratio.

The maximum local deviation of the observed data from the expected background is $1.4$ standard deviations, seen in the last bin of the $\bstar$ \textit{1L hadT} signal region (Figure~\ref{F:m_tot_only_hadTop}).
The value is computed using a moving-window algorithm~\cite{choudalakis2011hypothesis} that compares data to the improved background estimate in the signal regions.
Given the absence of a significant excess, limits are set on the production cross-section of both $\Bprime$ and $\bstar$ multiplied by the branching ratio of the decay to \Wt{}.

\begin{figure}[tbh]%
  \hfill%
\centering
  \subfigure[]{%
    \includegraphics[width=0.48\textwidth]{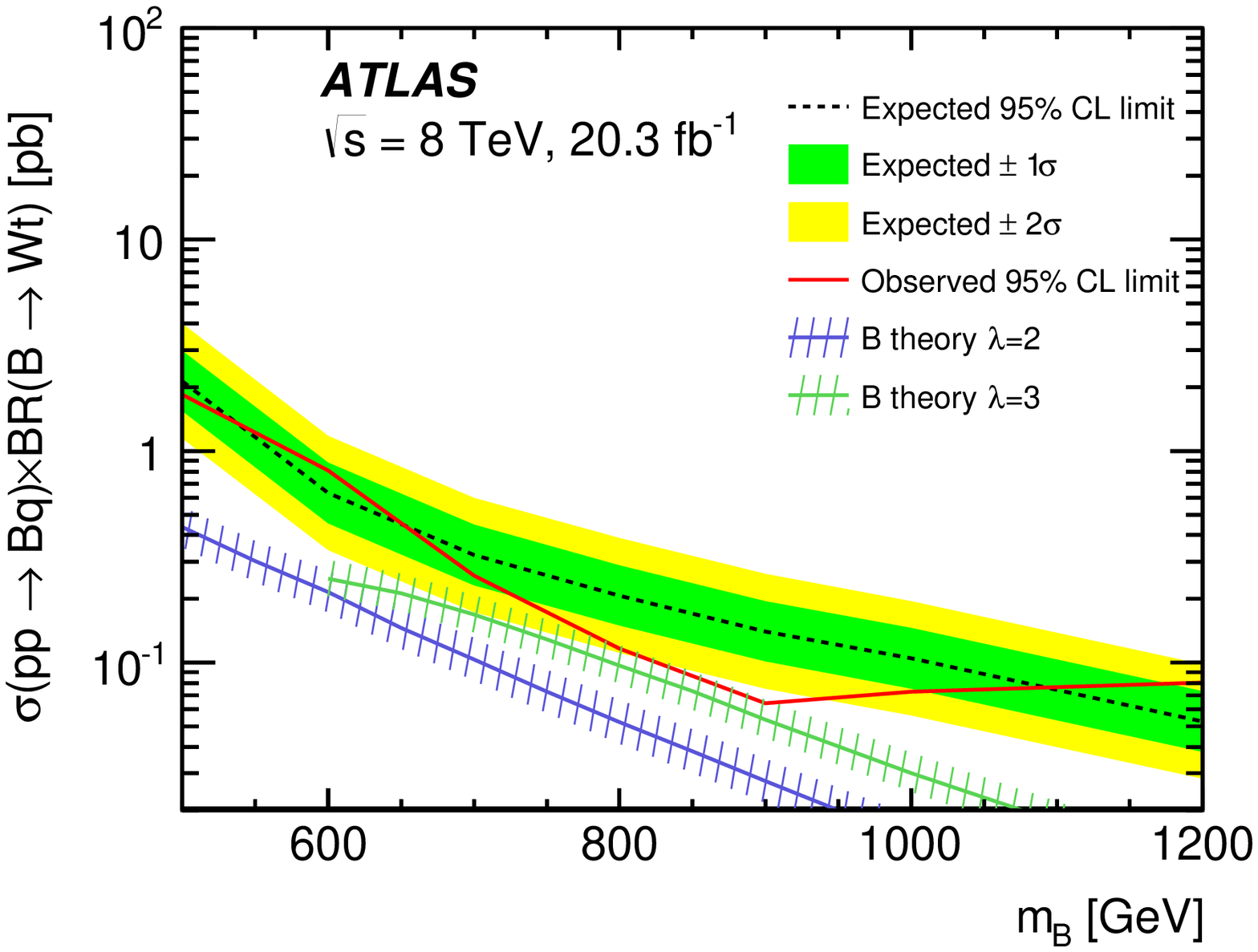}%
    \label{fig:bprime_xs_mass}%
  }%
  \hfill%
  \subfigure[]{%
    \includegraphics[width=0.48\textwidth]{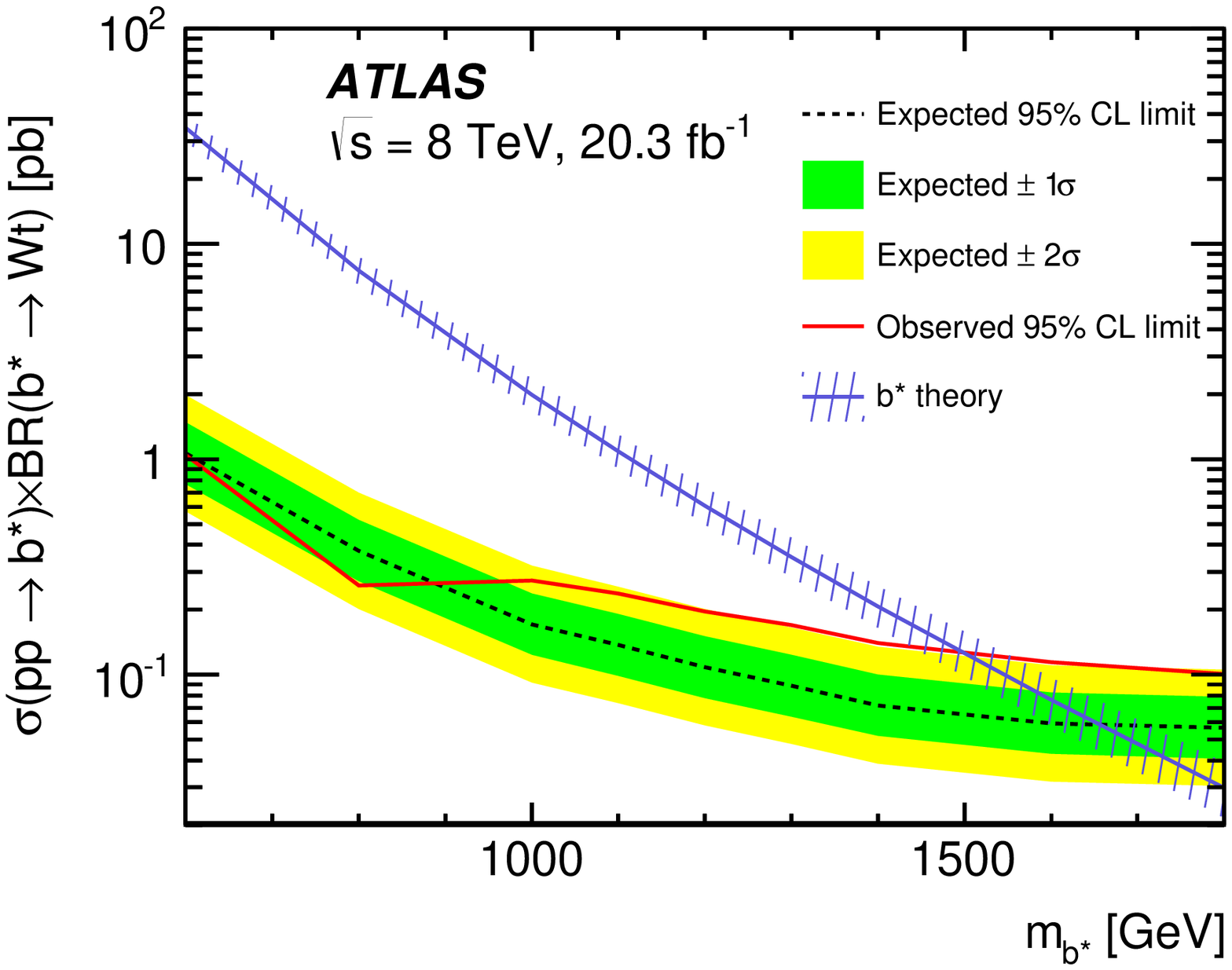}%
    \label{fig:bstar_xs_mass}%
  }%
  \hfill%
  \null%
  \caption{Expected (dashed black line) and observed (solid red line) $95\%$ CL limits on the \subref{fig:bprime_xs_mass} $\Bprime$ and the \subref{fig:bstar_xs_mass}
  $\bstar$ production cross-sections times branching fractions plotted against the mass.
  The uncertainty band around the expected limit indicates the variations by $\pm$ 1 and $\pm$ 2 standard deviations.
  The single-lepton and the dilepton channels are combined.
  Also shown are the theory predictions, which for $\Bprime$ are for two different coupling values $\lambda$ in the mass range where these are valid.  
  The couplings $\fg=\fL=\fR=1$ are assumed for the $\bstar$ theory predictions.
  For both the $\bstar$ and the $\Bprime$ signals, the uncertainty in the theory band includes the scale and the PDF uncertainties.}
  \label{fig:bxxx_xs_mass}%
\end{figure}%

The resulting expected and observed limits at the $95\%$ CL on cross-section times branching ratio are presented as functions of the $\Bprime$ (Figure~\ref{fig:bprime_xs_mass}) and $\bstar$ (Figure~\ref{fig:bstar_xs_mass}) mass, and are compared to the theory predictions.
No large deviations of the observed limits from the expected limits are found.
The observed limits for high masses are slightly worse than expected, because the corresponding fits are sensitive to the mild excess in the last bin of the signal region in the single-lepton channel (see above).

The observed (expected) mass limit is defined by the intersection of the theory line with the observed (expected) cross-section limit.
For the $\Bprime$ production, no limit on the mass and only limits on the cross-section times branching ratio can be set. 
For $\bstar$ production with couplings of $\fg=\fL=\fR=1$, the cross-section limits are translated into an observed (expected) lower limit on
the $\bstar$ mass of \SI{1500}{\GEV} (\SI{1660}{\GEV}). 

The cross-section is parameterised as a function of the couplings $\fg$ and $\fLR$ to extract the limits on these couplings for several mass hypotheses. 
The resulting contours are shown in Figure~\ref{fig:bxxx_xs_coupling}.

\begin{figure}[bth]
  \hfill%
\centering
  \subfigure[]{%
    \includegraphics[width=0.48\textwidth]{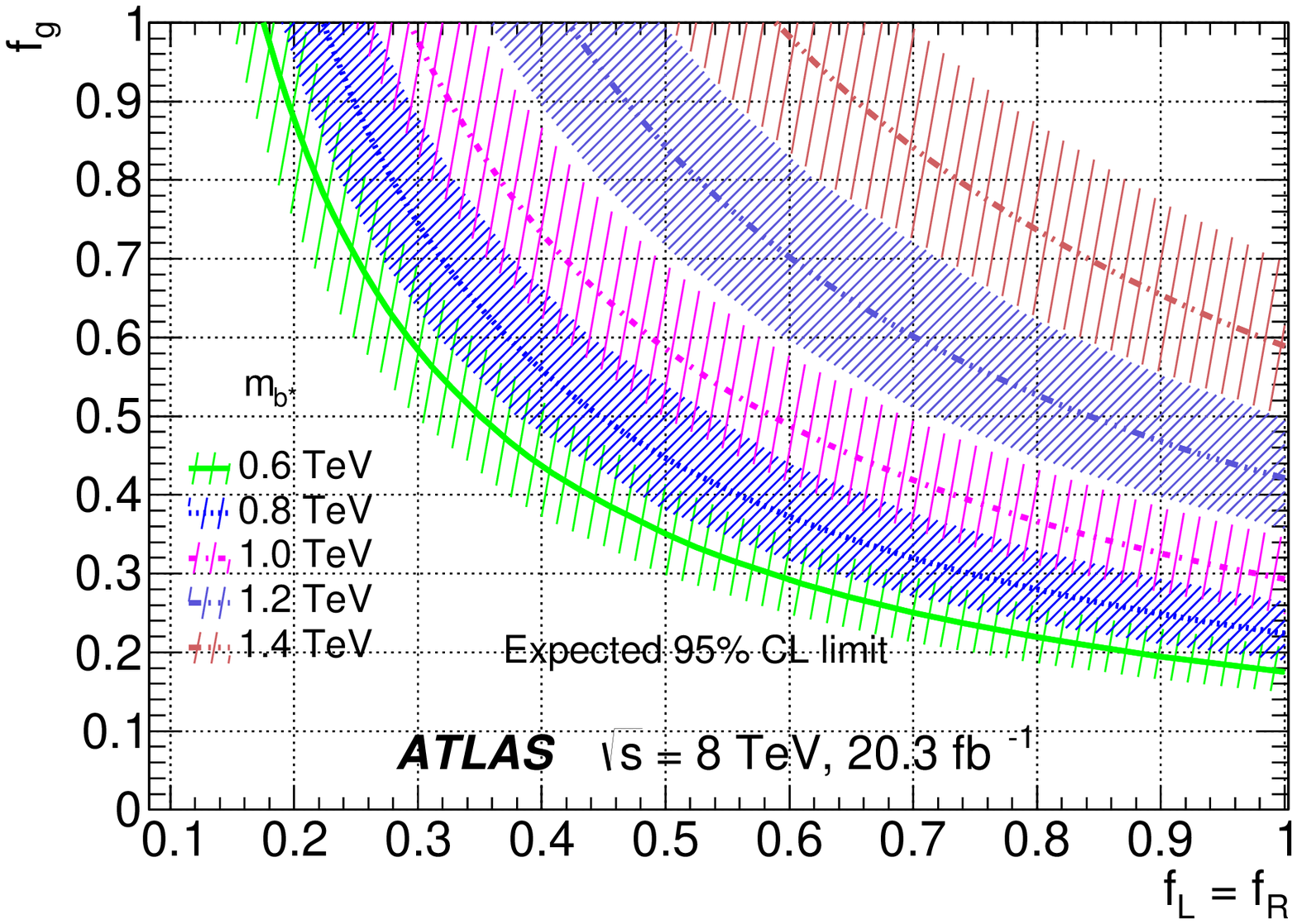}%
    \label{fig:bprime_xs_coupling}%
  }%
  \hfill%
  \subfigure[]{%
    \includegraphics[width=0.48\textwidth]{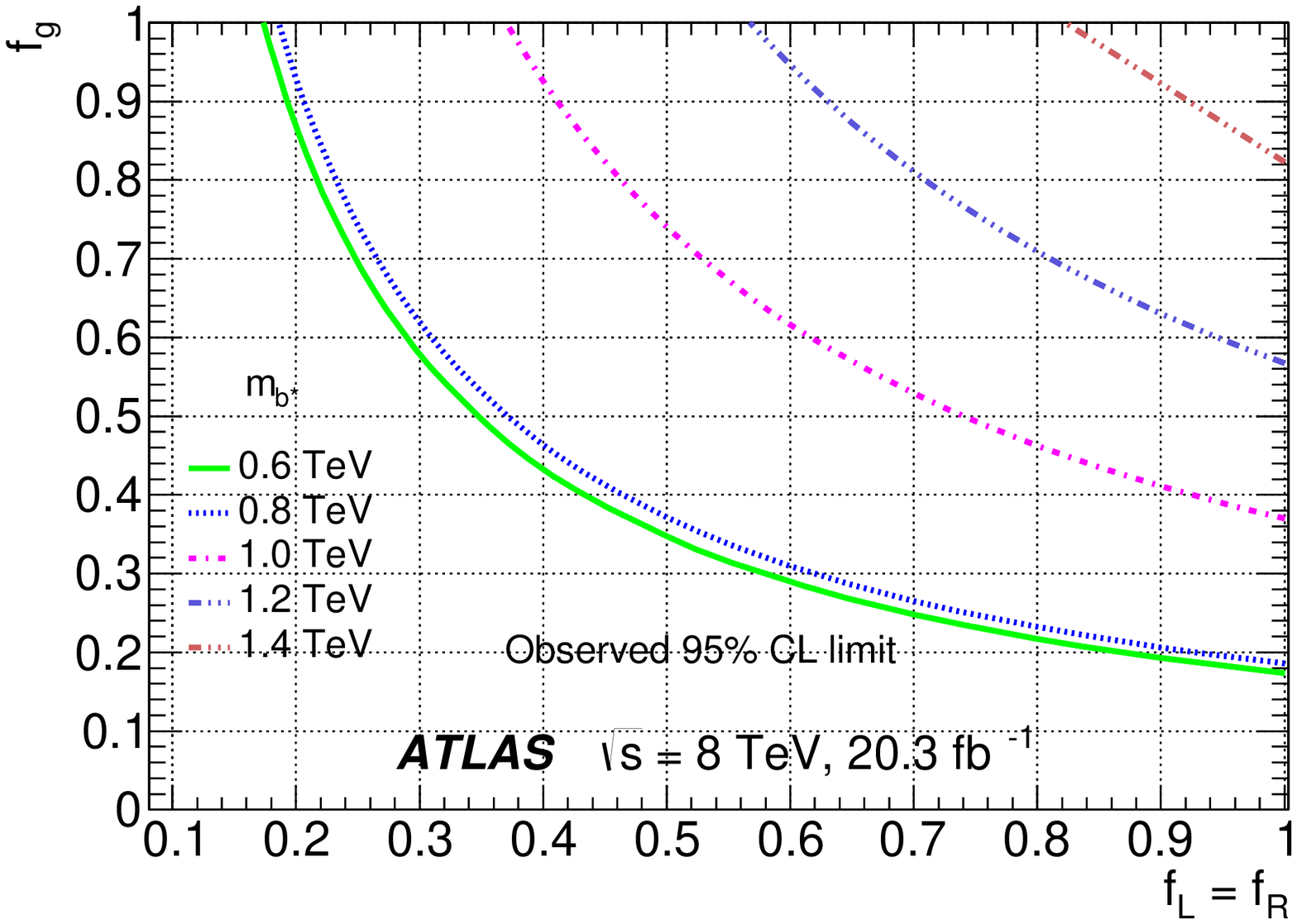}%
    \label{fig:bstar_xs_coupling}%
  }%
  \hfill%
  \null%
  \caption{Expected \subref{fig:bprime_xs_coupling} and observed \subref{fig:bstar_xs_coupling} limit contours at the $95\%$
  CL as a function of the coupling parameters for several choices of $\mbstar$. The surrounding shaded bands correspond to $\pm 1$ standard derivations
  around the expected limits. For a given mass, couplings above the corresponding contour line are excluded.
  }%
  \label{fig:bxxx_xs_coupling}%
\end{figure}

}%

\FloatBarrier
\section{Conclusion}
\label{sec:conclusion}

A search for the production of singly produced fourth-generation vector-like quarks and excited quarks with charge $\pm\frac{1}{3}e$ (down-type) in which the
\Bprime, produced via the weak interaction, and \bstar\, produced via the strong interaction with chromomagnetic couplings, decay into $\Wt$ has
been performed using \SI{20.3}{\per\fb} of $\sqrt{s}=\SI{8}{\TEV}$ proton--proton collision data delivered by the Large Hadron Collider and recorded with the ATLAS detector.
Final states with one or two leptons and with one, two or three jets are considered. 
The observations are consistent with the Standard Model expectations. Limits are set for \Bprime\ and \bstar\ models by combining the results from the single-lepton and dilepton channels. The \bstar\ models with $\fg=\fL=\fR=1$ and decay to $\Wt$ are excluded up to masses of \SI{1500}{\GEV} at 95\% CL. 
Upper limits on the production cross-sections times branching ratios for different vector-like \Bprime\ and excited \bstar\ quark masses are also provided.
The \bstar\ limit is an improvement over the previous ATLAS result and yields similar exclusion limits as the recent CMS result. 
The cross section limit on the single production of vector-like \Bprime\ quarks extends the set of existing \Bprime\ searches into the $\Wt$ final state using a novel approach for boosted event topologies.

\section*{Acknowledgements}

We thank CERN for the very successful operation of the LHC, as well as the
support staff from our institutions without whom ATLAS could not be
operated efficiently.

We acknowledge the support of ANPCyT, Argentina; YerPhI, Armenia; ARC, Australia; BMWFW and FWF, Austria; ANAS, Azerbaijan; SSTC, Belarus; CNPq and FAPESP, Brazil; NSERC, NRC and CFI, Canada; CERN; CONICYT, Chile; CAS, MOST and NSFC, China; COLCIENCIAS, Colombia; MSMT CR, MPO CR and VSC CR, Czech Republic; DNRF, DNSRC and Lundbeck Foundation, Denmark; IN2P3-CNRS, CEA-DSM/IRFU, France; GNSF, Georgia; BMBF, HGF, and MPG, Germany; GSRT, Greece; RGC, Hong Kong SAR, China; ISF, I-CORE and Benoziyo Center, Israel; INFN, Italy; MEXT and JSPS, Japan; CNRST, Morocco; FOM and NWO, Netherlands; RCN, Norway; MNiSW and NCN, Poland; FCT, Portugal; MNE/IFA, Romania; MES of Russia and NRC KI, Russian Federation; JINR; MESTD, Serbia; MSSR, Slovakia; ARRS and MIZ\v{S}, Slovenia; DST/NRF, South Africa; MINECO, Spain; SRC and Wallenberg Foundation, Sweden; SERI, SNSF and Cantons of Bern and Geneva, Switzerland; MOST, Taiwan; TAEK, Turkey; STFC, United Kingdom; DOE and NSF, United States of America. In addition, individual groups and members have received support from BCKDF, the Canada Council, CANARIE, CRC, Compute Canada, FQRNT, and the Ontario Innovation Trust, Canada; EPLANET, ERC, FP7, Horizon 2020 and Marie Skłodowska-Curie Actions, European Union; Investissements d'Avenir Labex and Idex, ANR, Region Auvergne and Fondation Partager le Savoir, France; DFG and AvH Foundation, Germany; Herakleitos, Thales and Aristeia programmes co-financed by EU-ESF and the Greek NSRF; BSF, GIF and Minerva, Israel; BRF, Norway; the Royal Society and Leverhulme Trust, United Kingdom.

The crucial computing support from all WLCG partners is acknowledged
gratefully, in particular from CERN and the ATLAS Tier-1 facilities at
TRIUMF (Canada), NDGF (Denmark, Norway, Sweden), CC-IN2P3 (France),
KIT/GridKA (Germany), INFN-CNAF (Italy), NL-T1 (Netherlands), PIC (Spain),
ASGC (Taiwan), RAL (UK) and BNL (USA) and in the Tier-2 facilities
worldwide.

\printbibliography

\newpage
\begin{flushleft}
{\Large The ATLAS Collaboration}

\bigskip

G.~Aad$^{\rm 85}$,
B.~Abbott$^{\rm 113}$,
J.~Abdallah$^{\rm 151}$,
O.~Abdinov$^{\rm 11}$,
R.~Aben$^{\rm 107}$,
M.~Abolins$^{\rm 90}$,
O.S.~AbouZeid$^{\rm 158}$,
H.~Abramowicz$^{\rm 153}$,
H.~Abreu$^{\rm 152}$,
R.~Abreu$^{\rm 116}$
% ...
\bigskip
\\
% ...
$^{11}$ Institute of Physics, Azerbaijan Academy of Sciences, Baku, Azerbaijan\\
% ...
$^{85}$ CPPM, Aix-Marseille Universit{\'e} and CNRS/IN2P3, Marseille, France\\
% ...
$^{90}$ Department of Physics and Astronomy, Michigan State University, East Lansing MI, United States of America\\
% ...
$^{107}$ Nikhef National Institute for Subatomic Physics and University of Amsterdam, Amsterdam, Netherlands\\
% ...
$^{113}$ Homer L. Dodge Department of Physics and Astronomy, University of Oklahoma, Norman OK, United States of America\\
% ...
$^{116}$ Center for High Energy Physics, University of Oregon, Eugene OR, United States of America\\
% ...
$^{151}$ Institute of Physics, Academia Sinica, Taipei, Taiwan\\
$^{152}$ Department of Physics, Technion: Israel Institute of Technology, Haifa, Israel\\
$^{153}$ Raymond and Beverly Sackler School of Physics and Astronomy, Tel Aviv University, Tel Aviv, Israel\\
% ...
$^{158}$ Department of Physics, University of Toronto, Toronto ON, Canada\\
% ...
\end{flushleft}


\clearpage

\end{document}